\documentclass{article}

\usepackage{graphicx,times} 
\usepackage{natbib}
\usepackage{float} 
\usepackage{amssymb,amsmath}
\usepackage{mathtools}
\usepackage[thmmarks,amsmath]{ntheorem} 
\usepackage[skip=2pt,textfont=rm,labelfont=rm]{caption}
\usepackage{bm} 
\usepackage{extarrows}
\usepackage[hang,flushmargin]{footmisc} 
\usepackage{mathrsfs} 
\usepackage{booktabs} 
\usepackage{tocloft}
\usepackage{diagbox}
\usepackage{threeparttable}
\usepackage{abbre}
\usepackage[a4paper=true,pagebackref=true]{hyperref}
\hypersetup{pdftitle = The title of my PDF, pdfauthor = My name, pdfsubject= The subject, pdfkeywords = keyword1 keyword2 keyword3} 
\hypersetup{colorlinks = true, linkcolor = green, anchorcolor = red, citecolor = blue, filecolor = red, urlcolor = red}

{
    \theoremstyle{nonumberplain}
    \theoremheaderfont{\bfseries}
    \theorembodyfont{\normalfont}
    \theoremsymbol{\mbox{$\Box$}}
    
}

\usepackage{theorem}

{
    \theoremheaderfont{\bfseries}
    \theorembodyfont{\normalfont}
    
}

\usepackage[a4paper]{geometry}                           
\begin{document}
\title{\bf Chemical evolution during the formation of molecular clouds} 
\date{}
\author{\sffamily Jingfei Sun$^{1,2}$, Fujun Du$^{1,2}$\\
    {\sffamily\small $^1$ Purple Mountain Observatory, Chinese Academy of Sciences, Nanjing 210023, China }\\
    \textbf{fjdu@pmo.ac.cn}\\
    {\sffamily\small $^2$ School of Astronomy and Space Science, University of Science and Technology of China, Hefei 230026, China }}
\renewcommand{\thefootnote}{\fnsymbol{footnote}}
\footnotetext[1]{Fujun Du: fjdu@pmo.ac.cn. }    
\maketitle

{\noindent\small{\bf Abstract:}
To study the chemical evolution during the formation of molecular clouds, we model three types of clouds with different density structures: collapsing spherical, collapsing ellipsoidal, and static spherical profiles.
The collapsing models are better than the static models in matching the observational characteristics in typical molecular clouds.
This is mainly because the gravity can speed up the formation of some important molecules (e.g., H$_2$, CO, OH) by increasing the number density during collapse.
The different morphologies of prolate, oblate, and spherical clouds lead to differences in chemical evolution, which are mainly due to their different evolution of number density.
We also study the effect of initial chemical compositions on chemical evolution, and find that H atoms can accelerate OH formation by two major reactions: O + H $\rightarrow$ OH in gas phase and on dust grain surfaces, leading to the models in which hydrogen is mainly atomic initially better match observations than the models in which hydrogen is mainly molecular initially.
Namely, to match observations, initially hydrogen must be mostly atomic.
The CO molecules are able to form even without the pre-existence of H$_2$.
We also study the influence of gas temperature, dust temperature, intensity of interstellar radiation field and cosmic-ray ionization rate on chemical evolution in static clouds.
The static CO clouds with high dust temperature, strong radiation field, and intensive cosmic rays are transient due to rapid CO destruction.
}

\vspace{1ex}
{\noindent\small{\bf Keywords:} ISM: abundances --- ISM: evolution --- ISM: molecules}

\section{Introduction}           
\label{sect:introduction}
It is known that atomic hydrogen can be converted into molecular hydrogen when dust grains or electrons act as a catalyst (\citealt{McCrea+McNally+1960, McDowell+1961}).
Almost all H$_2$ form on dust grain surfaces in the local universe where the metallicity is relatively high (\citealt{Ballesteros-Paredes+etal+2020}).
It is not yet clear how diffuse atomic gas turns into dense molecular gas in the Milky Way.
The primary mechanisms of molecular cloud formation which have been proposed are thermal instabilities (\citealt{Field+1965}), converging flows (\citealt{Bania+Lyon+1980}), stellar feedback process such as the expansion of HII regions and supernova blast wave, cloud-cloud collisions(\citealt{Oort+1954, Field+Saslaw+1965}), agglomeration of smaller clouds (\citealt{Casoli+Combes+1982}), gravitational instabilities (\citealt{Goldreich+Lynden-Bell+1965}), magnetic instabilities (\citealt{Parker+1966}) and magneto-rotational instabilities (\citealt{Balbus+Hawley+1991}).
Both observations and theories show the properties of molecular clouds depend on the galactic environment, so different mechanisms of molecular cloud formation may dominate different clouds.

There are two distinct chemical phases in the neutral clouds: atomic and molecular.
The atomic phase mainly consisted of neutral atomic hydrogen (HI) can be traced by the HI hyperfine transition at $\lambda = 21$ cm. 
Since the first detection of the 21 cm line by the Harvard team in 1951 (\citealt{Ewen+Purcell+1951}), there have been more and more studies on atomic hydrogen.
\citet{Li+Goldsmith+2003} and \citet{Goldsmith+Li+2005} used the HI narrow self-absorption (HINSA) technique to study atomic hydrogen in dark clouds. 
Furthermore, the transition from atomic to molecular hydrogen in dense clouds was studied by modeling the HINSA feature from cold atomic hydrogen (\citealt{Goldsmith+etal+2007}).
Numerous HI self-absorption observations (e.g., \citealt{Bigiel+etal+2008,Barriault+etal+2010,Stanimirovic+etal+2014,Burkhart+etal+2015}) indicate an upper column density threshold of $\sim10^{21}$ cm$^{-2}$ for cold HI in molecular clouds, consistent with semi-analytical models (\citealt{Sternberg+etal+2014,Bialy+Sternberg+2016}). 
The hydrogen molecule is non-polar and lacks a permanent electric dipole moment, hence it can only radiate through ro-vibrational or very weak quadrupole and induced dipole rotational transitions (\citealt{Field+etal+1966}).
Furthermore, the vibrational ground state with rotation level $J = 2$ is 512K above $J = 0$ and requires a temperature of hundreds of Kelvin to excite. These are reasons why the direct detection of H$_2$ is difficult by the emission of H$_2$ in cold interstellar medium.
The molecular phase is primarily traced by other molecular species, and one of the most common molecules to trace H$_2$ is the CO molecule, but CO as a tracer also has its shortcomings, one being that CO cannot trace all the H$_2$ gas.

The existence of molecular gas that cannot be traced by CO (CO-dark molecular gas) has been known for many years (\citealt{Lada+Blitz+1988}).
\cite{Tielens+Hollenbach+1985} predicted the existence of H$_2$ and C$^+$ layers from theoretical models of molecular clouds.
At the same time, various observations also indicate the existence of CO-dark molecular gas (\citealt{Reach+etal+1994, Grenier+etal+2005, Langer+etal+2010, PlanckCollaboration+etal+2011, Pineda+etal+2013, Langer+etal+2014}).
It has been explored whether OH cloud serve as a more effective tracer of CO-dark molecular gas (\citealt{Li+etal+2015}), so OH molecule is critical for studying interstellar medium.

There have been many observations of CO and OH molecules in molecular clouds.
\cite{Sofia+etal+2004} acquired a CO abundance of $\sim3.2{\times}10^{-4}$ toward some translucent sight lines in the galaxy.
In some CO-poor clouds, the CO abundance can be even lower than $\sim2.1\times 10^{-6}$(\citealt{Tang+etal+2017}).
These observations suggest there is a great difference in the CO abundance for different clouds.
Both radio and UV observations suggest that OH abundance is in a wide range.
\cite{Crutcher+1979} reviewed the data for diffuse clouds and concluded a OH abundance of $5{\times}10^{-8}$.
The OH abundance is close to $2{\times}10^{-8}$ by observations for a sample of compact extragalactic mm-wave continuum sources (\citealt{Liszt+Lucas+1996}).
\cite{Weselak+etal+2010} detected OH molecule toward 16 translucent sight lines and obtained a average OH abundance of $1.05{\times}10^{-7}$.
The OH abundance in molecular clouds ranges from $5.7{\times}10^{-8}$ to $4.8{\times}10^{-6}$ (\citealt{Tang+etal+2021}).

In this work we explore to what extent can we use chemical modeling to constrain their formation and early stages of clouds.
We use Chempl in this work, and we adopt a large reaction network.
We focus on the importance of some physical parameters for chemical models.
We model three types of density structures: gravitationally collapsing uniform spherical and ellipsoidal (prolate and oblate), and static spherical profiles.

This paper is organized as follows.
In Sect~\ref{sect:Meth}, we briefly describe the Chempl code.
Sect~\ref{sect:density and conditions} shows three types of density profiles, physical parameters, and initial abundances used in our models. 
In Sect~\ref{sect:results}, we present chemical evolution of four species (H, H$_2$, CO, and OH) in these clouds.
Sect~\ref{Sec:influence of physical parameters} show the influence of the gas temperature ($T_{\rm gas}$), the dust temperature ($T_{\rm dust}$), the intensity of interstellar radiation field ($G_0$, the ratio between the actual radiation field and \cite{Draine+1978} interstellar field), and cosmic-ray ionization rate ($\zeta$, H$_2$ ionization per sec) on abundances for these four species.
In Sect~\ref{sect:discussions}, we discuss the influence of initial abundances, density profiles, and physical parameters on the evolution of clouds.
Conclusions are presented in Sect~\ref{sect:conclusions}.
Appendix~\ref{Figures} gives additional figures related to main part of this work.

\section{Methodology}
\label{sect:Meth}
There are a few codes for chemical modeling, e.g., Nahoon (\citealt{Wakelam+etal+2012}), MAGICKAL (\citealt{Garrod+2013}), KROME (\citealt{Grassi+etal+2014}), Astrochem (\citealt{Maret+Bergin+2015}), Nautilus (\citealt{Ruaud+etal+2016}), UCLCHEM (\citealt{Holdship+etal+2017}), and Chempl (\citealt{Du+2021}).
The 1D and 2D models in this paper are carried out using the Chempl code (\citealt{Du+2021}), which works with comprehensive gas-grain chemical networks.
It supports the three-phase (gas, dust grain surface and mantle) formulation of interstellar gas-grain chemistry.
For the current work the UMIST 2012 network (\citealt{McElroy+etal+2013}) is used, augmented by grain surface reactions.  
In total the network contains 7721 reactions between 703 species.

\begin{table}
  \begin{center}
  \begin{minipage}[]{120mm}
  \caption[]{Some physical parameters and their typical values used in models\label{Tab1}}\end{minipage}
  \setlength{\tabcolsep}{5pt}
  \small
   \begin{tabular}{ll}
    \hline
    \hline\noalign{\smallskip}
  $G_0$ & 1\\
  $T_{\rm dust}$ & 15 K\\
  $T_{\rm gas}$ & 30 K\\
  $\zeta$ & $3{\times}10^{-17}$ s$^{-1}$\\
  Dust-to-gas mass ratio & 0.01\\
  Dust albedo & 0.6\\
  Dust grain radius & 0.1 $\mu$m\\
  Dust material density & 2.0 g cm$^{-3}$\\
  Dust site density & $10^{15}$ cm$^{-2}$\\
  Chemical desorption efficiency & 0.05\\
  Mean molecular weight & 1.4\\
    \hline\noalign{\smallskip}
  \end{tabular}
  \end{center}
\end{table}
Chemical modeling requires a lot of physical parameters, but here we only focus on a few physical parameters that are most relevant for chemical reaction rates.
Some important physical parameters used in models are listed in Table~\ref{Tab1}. 
At present the temperature profiles are input by hand instead of from self-consistent calculation based on cooling and heating balance.
It is necessary to provide a suitable value of $\zeta$ and $G_0$ for the models, because the destruction rates of most molecules are greatly affected by cosmic rays and ultra-violet (UV) photons.
The cosmic-ray ionization rate is taken to be a constant for simplicity, because it is only weakly attenuated in high column density regions (\citealt{Padovani+etal+2018}).
We adopt a common value of $\zeta$ and $G_0$ in Table~\ref{Tab1} for the majority of this paper, but also consider the influence of different values of $\zeta$ and $G_0$ on some chemical compositions in Sect~\ref{Sec:influence of physical parameters}.
The shielding effects of H$_2$ and CO greatly influence the chemical evolution of molecular clouds, so we take the self- and mutual-shielding of H$_2$ and CO molecules into account.
The self-shielding of H$_2$ from photodissociation is treated using the approach of \cite{Draine+Bertoldi+1996}, while the shielding of CO is implemented based on \citet{Morris+Jura+1983}.

\section{Density distribution and initial conditions}
\label{sect:density and conditions}
The gas density is a critical parameter for chemical evolution.
In this work we adopt three types of density profiles to study their differences in chemical evolution.
While the actual dynamic evolution may not follow a simple process described by analytical formulas, some qualitative insights can be gained through such a study.

\subsection{The gravitationally collapsing uniform spherical cloud}
Here we consider an isolated spherical cloud with uniform density under self-gravity, neglecting any rotation or magnetic field.
The cloud collapses toward the center under self-gravity, through a series of spherically symmetric states.
For its mathematical description, we follow the formulation of \citet{Lin+etal+1965}.
To maintain both the spherical shape and uniform density during the collapse, a particle with coordinates ($x_0$,$y_0$,$z_0$) at time $t$ = 0 must subsequently be at ($x$,$y$,$z$) with
\begin{equation}
  x = x_0X(t), ~~~~~~y = y_0Y(t), ~~~~~~z = z_0Z(t),
  \label{eq:E1}
\end{equation}
where $X(t),Y(t),Z(t)$ are dimensionless lengths, and the uniform number density is given by
\begin{equation}
  n(t) =  \frac{n_0}{X(t)Y(t)Z(t)}, 
  \label{eq:E2}
\end{equation}
where $n_0$ is the initial number density.

For the spherical case, we use spherical coordinates ($r,\theta, \varphi$), and then give the uniform number density:
\begin{equation}
  \label{eq:E4}
  n =  \frac{n_0}{R^3(t)},
\end{equation}
where the dimensionless radius is defined by $R(t) \equiv r/r_0$, and $r_0$ is initial radius.
The time evolution of a spherical cloud is given by \citet{Lin+etal+1965}
\begin{equation}
  \label{eq:E5}
  R = \cos^2\delta ,~~~ \tau = (\frac{8\pi}{3})^{-\frac{1}{2}}(\delta+\frac{1}{2}\sin2\delta),~~~ \tau \equiv (G\rho_0)^{\frac{1}{2}}t, 
\end{equation}
where $\tau$ is a dimensionless time (note that $(G\rho_0)^{1/2}$ is within one order of magnitude of the inverse of free-fall timescale), $G$ is the gravitational constant, $\rho_0$ is the initial mass density, and $\delta$ is an implicit function of time, which is needed to numerically calculate the evolution of density as a function of time.
From Eqs~\ref{eq:E4}-\ref{eq:E5}, we get
\begin{equation}
  \label{eq:E7}
  n(t) = \frac{n_0}{(\cos\delta(t))^6}.
\end{equation}

\subsection{The gravitationally collapsing uniform ellipsoidal cloud}
\label{sect:EllipsoidalCloudModel}
Next, we consider the cases of uniform oblate and prolate spheroids, using the formula of \cite{Lin+etal+1965}. 
For the spheroidal cases, we use cylindrical coordinates ($r,\theta,z$), and the uniform number density is
\begin{equation}
  \label{eq:E9}
  n =  \frac{n_0}{R^2(t)Z(t)},
\end{equation}
where the dimensionless radius is defined by $R(t) \equiv r/r_0$, $r_0$ is initial radius, the dimensionless height is defined by $Z(t) \equiv z/z_0$, and $z_0$ is initial height.

The time evolution of a oblate ellipsoidal cloud is given by \citet{Lin+etal+1965}
\begin{equation}
  \label{eq:E10}
  E = 1+E_2(\tan\delta)^2+E_4(\tan\delta)^4,~~~ a_0\tau = (\delta +\frac{1}{2}\sin2\delta)-(\frac{a_2}{a_0})(\delta - \frac{1}{2}\sin2\delta), ~~~ R = \cos^2\delta, ~~~ Z = ER,
\end{equation}
where $E_2,E_4,a_0,a_2$ are constants depending on the ellipticity, the dimensionless time $\tau$ is defined by Eq~\ref{eq:E5}, and $\delta$ is an implicit function of time.
From Eqs~\ref{eq:E9}-\ref{eq:E10}, we get 
\begin{equation}
  \label{eq:E11}
  n(t) = \frac{n_0}{E(\cos\delta(t))^6}.
\end{equation}

Similarly, the time evolution of a prolate ellipsoidal cloud is obtained by \citet{Lin+etal+1965}
\begin{equation}
  \label{eq:E12}
  E = 1+E_2(\tan\delta)^2+E_4(\tan\delta)^4, ~~~ a_0\tau = (\delta +\frac{1}{2}\sin2\delta)-(\frac{a_2}{a_0})(\delta - \frac{1}{2}\sin2\delta), ~~~Z = \cos^2\delta,  ~~~R = EZ.
\end{equation}
Then, we get the number density vary with time as
\begin{equation}
  \label{eq:E13}
  n(t) = \frac{n_0}{E^2(\cos\delta(t))^6}.
\end{equation}

Note that the first two expressions in Eq~\ref{eq:E10} and \ref{eq:E12} are just the first few terms of a series expansion instead of the exact solution. 
For a specific shape and time, we can calculate the value of $\delta$ and $E$, and then acquire the number density for an ellipsoidal cloud.
Here the collapse timescale is defined as the time when the semiminor axis of an ellipsoid shrinks to zero, which is used as the cloud evolution time.  
The value of $E$ is zero at the time, so we can get the collapse timescale by the value of $\delta$.
When the ellipticity of an ellipsoid is close to zero, its collapse timescale becomes the same as the free-fall timescale for the spherical case.

\begin{figure} 
  \centering
  \includegraphics[width=0.9\textwidth, angle=0]{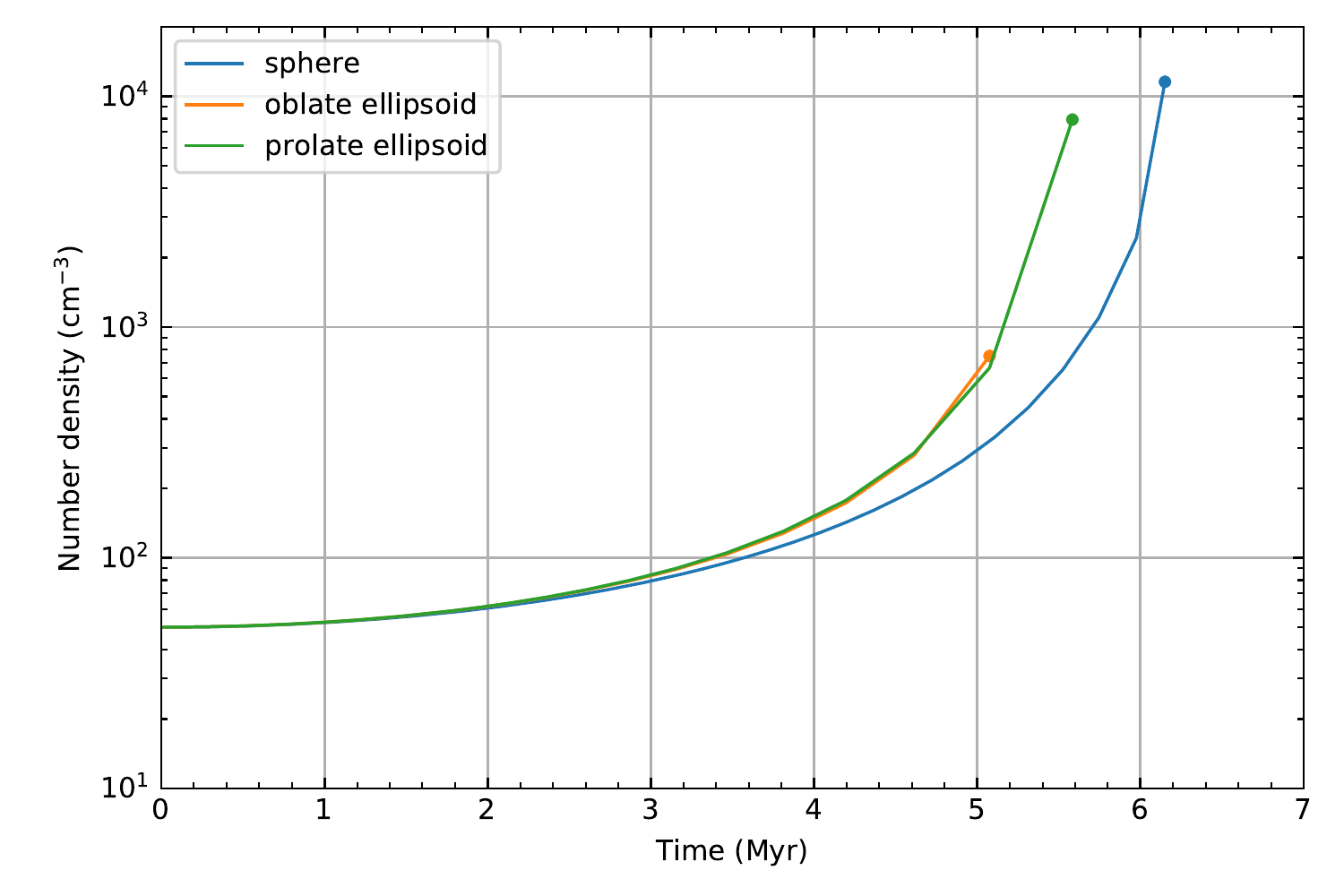}
  \caption{The number density as a function of time in different collapsing models (sphere, oblate ellipsoid, and prolate ellipsoid). Note that their end time points are determined by the morphology, initial density, and final semiminor axis. } 
  \label{NumberDensity}
\end{figure}
To see the differences in the number density between different collapsing models (sphere, oblate ellipsoid, and prolate ellipsoid) with identical initial number density ($n_0 = 50$ cm$^{-3}$), we show the number density as a function of time in Fig~\ref{NumberDensity} (initial shape and number density are introduced in Sect~\ref{sect:results}).
When these semiminor axises of collapsing clouds shrink to typical scales of molecular clouds (1-3 pc), we artificially cease their dynamical evolutions, then obtain their end time points.
Note the prolate ellipsoid is larger than the oblate ellipsoid initially (given in Sect~\ref{sect:results}), so we adopt a large final scale for the prolate one.
The collapse of the sphere is slower than the oblate or prolate spheroid, so morphological anisotropy can lead to large differences in the evolution of number density. 
At late stages, the shrinkage of the prolate and oblate ellipsoids are mainly concentrated on their short axis, which leads to a more rapid change in volume and density for the prolate one.
Their final number densities are slightly artificial due to the range of typical scales, but it does not change the conclusion that the prolate one has a higher final number density.
The oblate spheroid approaches a thin disk at final time, while the prolate spheroid becomes a thin cylinder.
In general, the dynamics for collapsing clouds with different shapes have a great difference at late stages.

\subsection{The static spherical cloud}
\label{sect:StaticModel}
The solution of self-similar collapse of isothermal spheres show that an $r^{-2}$ law holds for the density distribution in the static or nearly static outer envelope (e.g., \citealt{Bodenheimer+Sweigart+1968, Shu+1977}).
Since the central density is infinite with this density distribution, we adopt another treatment as \cite{Lee+etal+1996}, where the total hydrogen number density $n_{\rm H}$ = $n$(H)+2$n$(H$_2$) as a function of radius varies as
\begin{equation}
  \label{eq:E14}
  n_{\rm H} = \frac{n_{\rm H, max}}{(1+\frac{c\cdot r}{r_{\rm max}})^2},
\end{equation}
where $n_{\rm H, max}$ is the hydrogen density at the cloud center ($r=0$), $r_{\rm max}$ is the cloud radius, and $c$ is a constant.
If the number density at the cloud surface $n_{\rm {H,min}}$ is known, we can get the value of the constant $c$ as 
\begin{equation}
  \label{eq:E15}
  c = \left(\frac{n_\text{H,max}}{n_\text{H,min}}\right)^{1/2}-1.
\end{equation} 

\subsection{Physical parameters}
\label{physical conditions}
We assume these clouds are exposed to the \cite{Draine+1978} interstellar radiation field (equal to $\sim$1.7 times the \cite{Habing+1968} local interstellar filed), which may be multiplied by an intensity scaling factor $G_0$ to represent a different radiation.
The interstellar medium near an early-type star is exposed to strong UV radiation filed, so we investigate the influence of strong radiation field on H, H$_2$, CO and OH abundances in Sect~\ref{subsect:G0}.
The cosmic-ray ionization rate is adopted as $3\times10^{-17}$ s$^{-1}$ for the majority of this paper (\citealt{Clark+etal+2019}).
Observations for some objects under different interstellar environments, using tracers such as H$_3^{+}$, indicate that there is a wide range of cosmic-ray ionization rate, from a few times $10^{-18}$ to a few times $10^{-14}$ s$^{-1}$ (e.g., \citealt{Hartquist+etal+1978, vanderTak+vanDishoeck+2000, Shaw+etal+2008, Indriolo+etal+2010, Indriolo+McCall+2012, Indriolo+etal+2015}). 
Thus we also study the influence of different cosmic-ray ionization rates on H, H$_2$, CO and OH abundances in Sect~\ref{subsect:CRIR}.
We assume dust is characterized by a representative grain radius $a$ = 0.1 $\mu$m and an average albedo of 0.6. 
The material density of dust is taken to be 2 g cm$^{-2}$ and the dust-to-gas mass ratio is taken to be 0.01. 
The surface site density is taken to be 1$\times10^{15}$ cm$^{-2}$, and the chemical desorption efficiency is taken to be 0.05 (\citealt{Du+2021}).
The main energy coupling of gas and dust is through gas-grain collisions. 
\cite{Bergin+Tafalla+2007} reviewed that the gas and dust tend to couple thermally via frequent collisions at high densities ($> 10^4$ cm$^{-3}$), and their temperatures are expected to converge.
Furthermore, the gas and dust in low density regions are thermally decoupled due to inadequate collisions, and the gas becomes warmer than the dust due to photoelectric heating.
In a few hundred number density photodissociation region, both observations and theoretical calculations indicate the gas temperature is about 30 K(\citealt{Wolfire+etal+2003,Tielens+2010}), and that dust temperature is close to a value of 15 K (\citealt{Ward-Thompson+etal+2002,Bianchi+etal+2003,Tielens+2010}), so we adopt these values in this work.
We also study the influence of different gas and dust temperature on H, H$_2$, CO, and OH abundances in Sect~\ref{Sec:influence of physical parameters}.
These physical parameters are summarized in Table~\ref{Tab1}.

Since these clouds that we study are spherical, a weighting factor for each shell should be applied to calculate the column density for specie $X$, which boosts the importance of outer shells and lower the importance of inner shells.
We define a weighted column density $N(X)$ for specie $X$ as (\citealt{Lee+etal+1996}) 
\begin{equation}
  \label{eq:E20}
  N(X) = A \sum_{i}\left[n_i(X)dL_i\left(\frac{r_i}{r_{\rm max}}\right)^2\right],
\end{equation}
where $A$ is a normalization constant, $n_i(X)$ is the number density for specie $X$ in shell $i$, $dL_i$ is the width of shell $i$, $r_i$ is the distance from the center for shell $i$, and $r_{\rm max}$ is the cloud radius.
The relation between the visual extinction $A_{\rm V}$ and the total column density of protons $N_{\rm H}$ is $A_{\rm V} = 5.34{\times}10^{-22}N_{\rm H}$ (\citealt{Bohlin+etal+1978}).
To study the influence of different clouds and initial abundances, we define the weighted column density for specie $X$ relative to total column density as the average abundance $f(X)$ = $N({X})/(2N({\rm H}_2)+N({\rm H}))$.

\subsection{Initial abundances}
\label{sub:initial abundance}
\begin{table}
  \begin{center}
  \begin{minipage}[]{100mm}
  \caption[]{Initial abundances with respect to total hydrogen\label{Tab2}}\end{minipage}
  \setlength{\tabcolsep}{10pt}
  \small
  \begin{threeparttable}
  \begin{tabular}{ll|ll|ll}
    \hline
    \hline
    \multicolumn{2}{c|}{First type / H} & \multicolumn{2}{c|}{Second type / H/H$_2$} & \multicolumn{2}{c}{Third type / H$_2$}\\
    \hline
  Species & Abundance & Species & Abundance & Species & Abundance\\
    \hline
  H & $-2{\times}10^{-5}{+}1$ & H & variable & H & 0.0\\
  H$_2$ & $1{\times}10^{-5}$ & H$_2$ & variable & H$_2$ & $5{\times}10^{-1}$\\
  He & $9{\times}10^{-2}$ & He & $9{\times}10^{-2}$ & He & $9{\times}10^{-2}$\\
  C$^+$ & $1.4{\times}10^{-4}$ & C$^+$ & $1.4{\times}10^{-4}$ & C$^+$ & $1.4{\times}10^{-4}$\\
  N & $7.5{\times}10^{-5}$ & N & $7.5{\times}10^{-5}$ & N & $7.5{\times}10^{-5}$\\
  O & $3.2{\times}10^{-4}$ & O & $3.2{\times}10^{-4}$ & O & $3.2{\times}10^{-4}$\\
  S$^+$ & $8{\times}10^{-8}$ & S$^+$ & $8{\times}10^{-8}$ & S$^+$ & $8{\times}10^{-8}$\\
  Si$^+$ & $8{\times}10^{-9}$ & Si$^+$ & $8{\times}10^{-9}$ & Si$^+$ & $8{\times}10^{-9}$\\
  Fe$^+$ & $3{\times}10^{-9}$ & Fe$^+$ & $3{\times}10^{-9}$ & Fe$^+$ & $3{\times}10^{-9}$\\
  Mg$^+$ & $7{\times}10^{-9}$ & Mg$^+$ & $7{\times}10^{-9}$ & Mg$^+$ & $7{\times}10^{-9}$\\
  Cl$^+$ & $4{\times}10^{-9}$ & Cl$^+$ & $4{\times}10^{-9}$ & Cl$^+$ & $4{\times}10^{-9}$\\
  Na$^+$ & $2{\times}10^{-9}$ & Na$^+$ & $2{\times}10^{-9}$ & Na$^+$ & $2{\times}10^{-9}$\\
  P$^+$ & $3{\times}10^{-9}$ & P$^+$ & $3{\times}10^{-9}$ & P$^+$ & $3{\times}10^{-9}$\\
  F & $4{\times}10^{-9}$ & F & $4{\times}10^{-9}$ & F & $4{\times}10^{-9}$\\
    \hline\noalign{\smallskip}
  \end{tabular}
  \begin{tablenotes}
    \item H, H/H$_2$, and H$_2$ represent respectively the model in which hydrogen is mainly atomic, shell-dependent H/H$_2$ ratio, molecular initially in all shells.
  \end{tablenotes}
  \end{threeparttable}
  \end{center}
\end{table}
To figure out the effect of the initial chemical compositions on the chemical evolution, we select three types of initial chemical compositions listed in Table~\ref{Tab2} (\citealt{McElroy+etal+2013, Du+2021}).
In the first type, we assume hydrogen is completely atomic initially for all shells. 
In the second type, the initial abundances except H and H$_2$ are the same as in first type.
For the initial abundances of H and H$_2$, we run our code with the first type initial conditions in all shells for a certain amount of time, and use the final abundances of H and H$_2$ as their initial abundances.
For these static clouds, we adopt the H/H$_2$ abundances at a time of $1{\times}10^6$ yr as their initial abundances in the second type.
For these collapsing spherical or ellipsoidal clouds, we adopt the H/H$_2$ abundances at $t = 6.15{\times}10^6$ yr (equal to collapsing timescale of a uniform spherical cloud) as their initial abundances.
In the third type, we start with molecular hydrogen in all shells.
Based on cloud types and initial chemical compositions, in total, 12 models listed in Table~\ref{Tab3} are constructed, which are labeled Model 1--12.
\begin{table}
  \begin{center}
  \caption[]{Cloud types and initial H/H$_2$ abundances in different models\label{Tab3}}
  \setlength{\tabcolsep}{10pt}
  \small
  \begin{tabular}{|c|c|c|c|}
  \hline
  \diagbox[innerwidth=7.5cm]{Cloud}{Model}{Initial hydrogen} & H & H/H$_2$ & H$_2$\\
  \hline
  Gravitationally collapsing uniform spherical cloud & Model 1 & Model 2 & Model 3\\
  \hline
  Gravitationally collapsing uniform oblate ellipsoidal cloud & Model 4 & Model 5 & Model 6\\
  \hline
  Gravitationally collapsing uniform prolate ellipsoidal cloud & Model 7 & Model 8 & Model 9\\
  \hline
  Static spherical cloud & Model 10 & Model 11 & Model 12\\
  \hline\noalign{\smallskip}
  \end{tabular}
  \end{center}
\end{table}
\section{Results}
\label{sect:results}
The H atoms in gas phase are reduced by adsorbing on the dust grain surfaces.
The formation of H$_2$ molecules on dust grain surfaces is treated as a normal two-body surface reaction: gH + gH $\rightarrow$ gH$_2$ in our models, which is different from some of previous works (as \citealt{Li+Goldsmith+2003,Holdship+etal+2017}).
The CO molecules are mainly produced in gas phase, and removed from the gas through photodissociation, cosmic-ray-induced photodissociation and adsorption reactions.
The OH molecules for different time are dominated by different reactions. 
Studying their abundances is helpful in constraining the initial compositions and evolution history of clouds.
\begin{figure} 
  \centering
  \includegraphics[width=1.0\textwidth, angle=0]{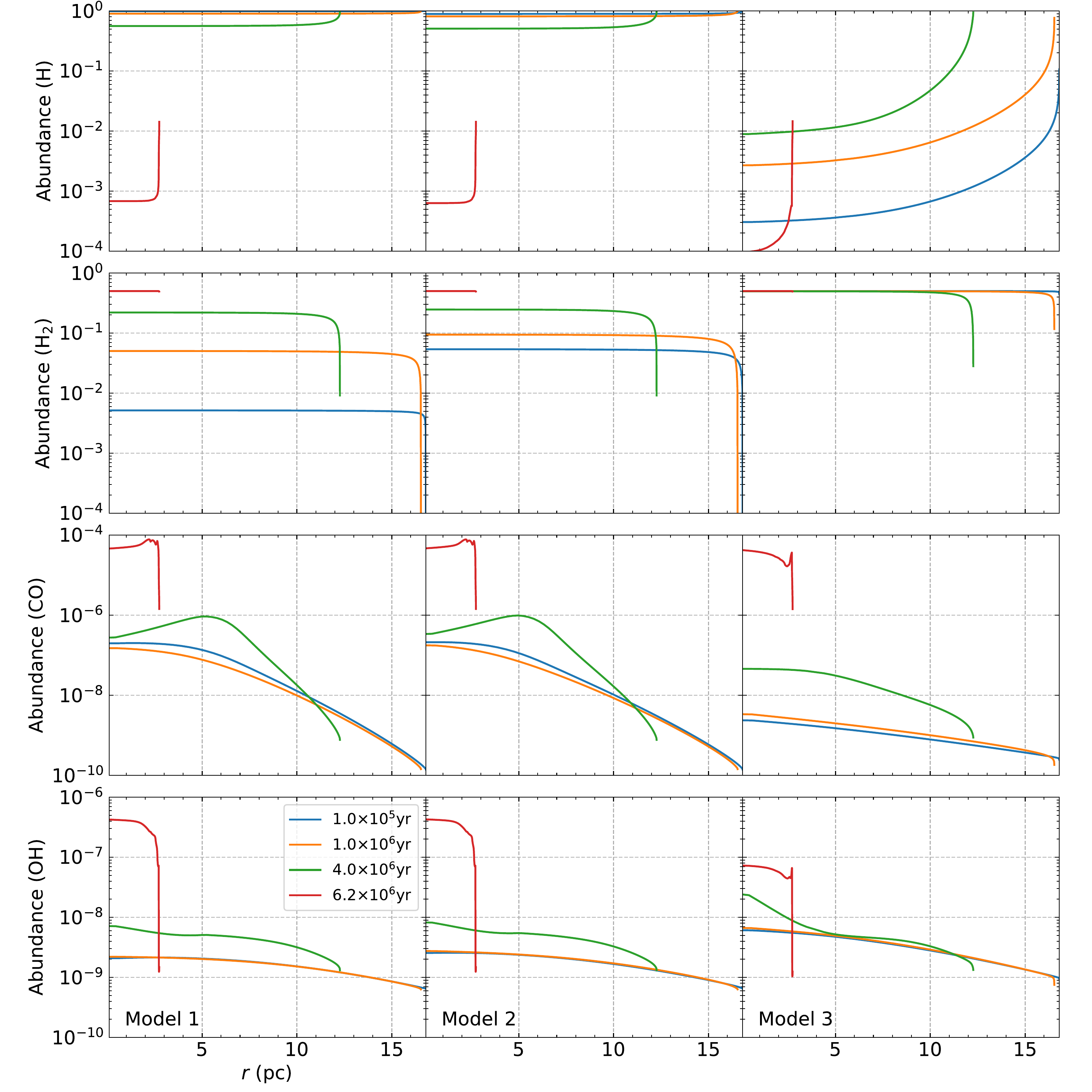}
  \caption{The abundances of H, H$_2$, CO, and OH as a function of radius $r$ in a gravitationally collapsing uniform spherical cloud at various time using models 1-3.
  Note that as time goes on, the cloud shrinks, and the ends of the curves move toward inner radius.} 
  \label{FigAbundanceForSphere}
\end{figure}

\subsection{The gravitationally collapsing uniform spherical cloud}
Initially the cloud has a radius of 16.8 pc and total mass of $3.4{\times}10^4$ M$_\odot$, which are selected so that the mass and spatial and dynamical scale mimic those of observed molecular clouds.
The initial number density of the cloud is 50 cm$^{-3}$, and the cloud collapse timescale is $6.2{\times}10^6$ yr.
The cloud radius changes from 16.8 to 2.7 pc, when the number density changes from 50 to $1.2{\times}10^4$ cm$^{-3}$. 
The total column density increases by a factor of about 2 for the spherical cloud.

Fig~\ref{FigAbundanceForSphere} shows the abundance profiles for H, H$_2$, CO, and OH at different time in models 1-3.
Since the results in Model 1 and 2 are similar, we discuss them in Model 1 as an example.
The cloud radius begins to change appreciably at $t\sim4{\times}10^6$ yr, leading to the increasing number density and boosted reaction rates in models 1-3.
In Model 1, the H$_2$ abundance for $t\sim6{\times}10^6$ yr is characterized by an essentially flat profile, with abundance $\sim0.5$, when the H abundance for all radius is low, being more abundant outwards. 
At the same time, the abundance of CO and OH for inner radius respectively reaches $4{\times}10^{-5}$ and $4{\times}10^{-7}$, while their abundances in outer layers are still extremely low due to intensive UV photons. 
In Model 3, the OH abundance for inner radius is lower than in Model 1 at late time.
This is mainly because H atoms can accelerate OH formation by two major reactions: O + H $\rightarrow$ OH in gas phase and on dust grain surfaces.
In the absence of H atoms, OH molecules are slowly produced by the reaction: O + H$_2$S$^+$ $\rightarrow$ HS$^+$ + OH.
The central HI density at late time is close to 1 cm$^{-3}$, which is in reasonable agreement with the value in \cite{Li+Goldsmith+2003}.
The H atoms on dust grain surfaces can effectively form H$_2$ molecules before desorption, so the H$_2$ formation efficiency is close to unity \citep{Li+Goldsmith+2003}. 
Note that the central HI density at a high dust temperature is clearly higher than 1 cm$^{-3}$ due to low H$_2$ formation efficiency, and we detailed discuss the reason in Sect~\ref{subsect:Tdust}.

\begin{figure} 
  \centering
  \includegraphics[width=14.0cm, angle=0]{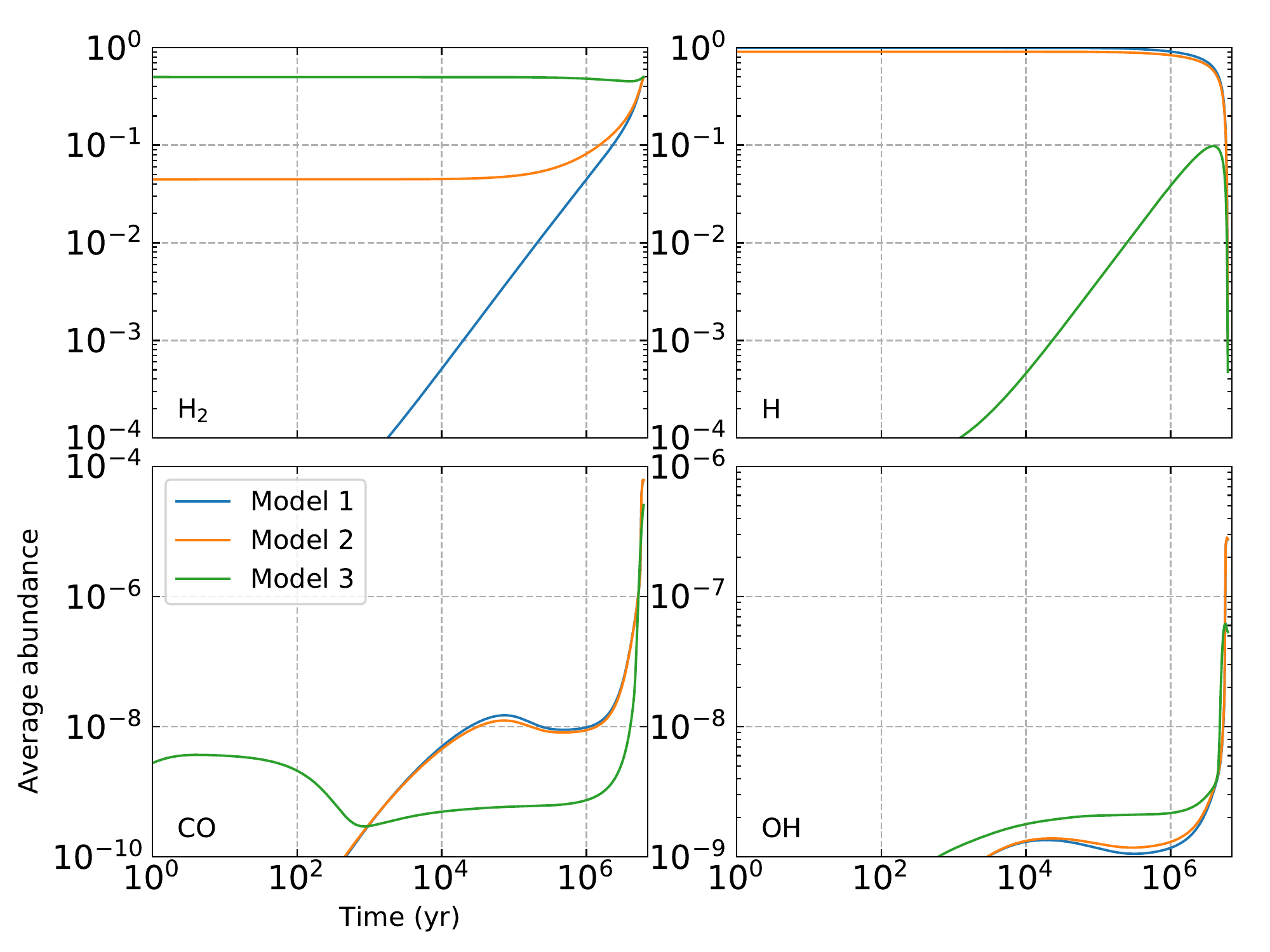}
  \caption{The average abundances for H, H$_2$, CO, and OH as a function of time using models 1-3.
   } 
  \label{FigAverageAbundanceForSphere}
\end{figure}
Fig~\ref{FigAverageAbundanceForSphere} shows the average abundances for above four species varying with time in models 1-3.
For models 1-3, the collapsing clouds at early time are the pre-CO phases, when CO has not formed in sufficient quantity to be detectable, and OH molecules are also difficult to detect. 
The average abundance of CO and OH in models 1-2 respectively grows to $5{\times}10^{-5}$ and $3{\times}10^{-7}$, which is close to their typical values in molecular clouds, when the cloud radius shrinks to $\sim3$ pc. 
At this time, the HI column density is close to $10^{20}$ cm$^{-2}$, which is identical with its observed value in \cite{Wannier+etal+1983}.
Accordingly, the collapsing spherical clouds in which hydrogen is mainly atomic initially can match the observational characteristics of molecular clouds.  
The OH abundance in Model 3 is lower than its minimum value of $5.7{\times}10^{-8}$ in molecular clouds from \cite{Tang+etal+2021}, which suggests the collapsing cloud in which hydrogen is molecular initially cannot match OH abundance in typical molecular clouds.

\subsection{The gravitationally collapsing uniform ellipsoidal cloud}
Oblate or prolate ellipsoidal clouds are axisymmetric, so we only discuss the abundance distributions for H, H$_2$, CO and OH in two-dimensional space.
The oblate ellipsoidal cloud has an initial number density of 50 cm$^{-3}$ and total mass of $1.4{\times}10^4$ M$_\odot$.
The semimajor axis is 16.8 pc and semiminor axis is 6.7 pc.
The cloud collapse timescale is $5.4{\times}10^6$ yr. 
Here we mainly show the variation of physical parameters along the short axis, which are mainly determined by the shortest distance from cloud surface.
At the final stage of our modeling, the semiminor axis shrinks to 0.9 pc, and the number density is about 750 cm$^{-3}$.
The column density increases by a factor of about 2 for the oblate ellipsoidal cloud.
We also model a prolate ellipsoidal cloud, using another set of physical parameters.
These physical parameters for oblate and prolate ellipsoidal clouds are listed in Table~\ref{Tab4}.

\begin{table}
  \begin{center}
  \begin{minipage}[]{100mm}
  \caption[]{Physical parameters for oblate and prolate ellipsoidal clouds \label{Tab4}}\end{minipage}
  \setlength{\tabcolsep}{10pt}
  \small
  \begin{tabular}{lcc}
    \hline
    \hline
       & Oblate ellipsoidal cloud & Prolate ellipsoidal cloud\\
    \hline
  Initial number density (cm$^{-3}$) & 50 & 50\\
  Mass (M$_\odot$) & $1.4{\times}10^4$ & $6.9{\times}10^4$\\
  Initial semimajor (pc) & 16.8 & 33.6\\
  Initial semiminor axis (pc) & 6.7 & 16.8\\
  Collapse timescale (yr) & $5.4{\times}10^6$ & $5.8{\times}10^6$\\
  Final semiminor axis (pc) & 0.9 & 1.6\\
  Final number density (cm$^{-3}$) & 750 & 7928\\
    \hline\noalign{\smallskip}
  \end{tabular}
  \end{center}
\end{table}
To save space, we put the abundance maps for H, H$_2$, CO and OH to Appendix~\ref{Figures}.
Figs~\ref{FigAtomicHydrogenAbundanceForOblateCase}-\ref{FigOHAbundanceForProlateCase} show their abundance distributions at different time in models 4-9.
The H$_2$ abundance in models 7-8 is more abundant than in models 4-5 at late times, suggesting the transition from H to H$_2$ is fast in high density $>750$ cm$^{-3}$ regions.
For $t=5.4{\times}10^6$ yr, H$_2$ molecules in Model 7 have been produced in large quantity, so the cloud is mainly composed of molecular hydrogen.
At the final stage, CO and OH molecules for models 4-5 are not formed in large quantity, while for models 7-8 their abundances reach a high level due to high number density.
The behaviors for these four species in Model 6 are different from in Model 9.
At late time, the abundances for H$_2$, CO and OH in Model 9 are more abundant than in Model 6, suggesting higher chemical reaction rates in Model 9.
In general, there are some differences in abundances of some species (e.g., OH, CO) between the oblate and prolate ellipsoid, which are mainly due to their different evolution of number density (see Sect~\ref{sect:EllipsoidalCloudModel}).
Namely, differences in clouds shape can lead to their differences in chemical evolution via the number density.

\subsection{The static spherical cloud}
The number density of the cloud as a function of radius is described in Sect~\ref{sect:StaticModel}.
Here, we take the number density at the cloud center $n_{\rm {H, max}}$ = $1{\times}10^4$ cm$^{-3}$ and the number density at the cloud surface $n_{\rm {H, min}}$ = $1{\times}10^2$ cm$^{-3}$.
The static cloud has a total radius of 7 pc, total visual extinction $A_{\rm V}$ of 10.4 mag, and total mass of $1.1{\times}10^4$ M$_{\odot}$.
Some observations for Class~I clouds have shown their lifetime is limited to a few Myr, which represent the duration of the `inert' CO phase without massive star formation (e.g., \citealt{Fukui+etal+1999,Kawamura+etal+2009}).
Hence the chemical modeling time is taken to be $8{\times}10^6$ yr in these static models.

\begin{figure} 
  \centering
  \includegraphics[width=1.0\textwidth, angle=0]{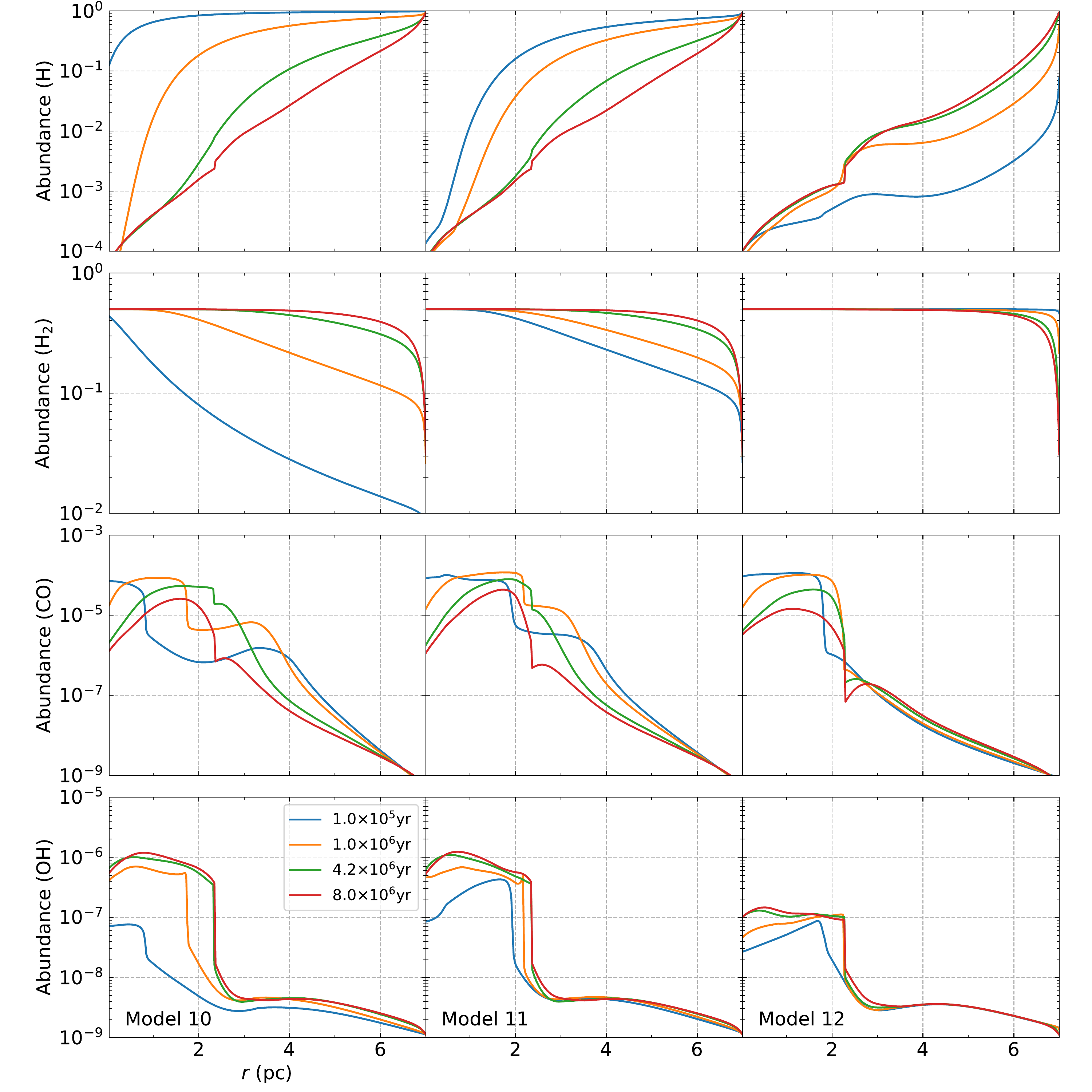}
  \caption{The abundances of H, H$_2$, CO and OH as a function of radius $r$ in a static spherical cloud at various time using models 10-12.
   } 
  \label{FigAbundanceForStatic}
\end{figure}
Fig~\ref{FigAbundanceForStatic} shows the abundance profiles for H, H$_2$, CO, and OH at different time in models 10-12.
The CO abundance for inner radius begins to decrease due to adsorption at $t\sim10^6$ yr in these static models, which is different from some works leaving CO adsorption out of account (e.g., \citealt{Lee+etal+1996}). 
At late time, the OH abundance of $\sim10^{-6}$ for inner radius in Model 10 is clearly higher than its value of $\sim10^{-7}$ in Model 12, which resembles the collapsing models.
There is a noteworthy rapid change for H, CO, and OH abundances at $r\sim2.3$ pc in these static models, corresponding to a visual extinction of $\sim2.0$ mag.
Our preliminary investigation suggests that the rapid change may be related to bistable solutions (low-ionization phase and high-ionization phase) in the nonlinear kinetic equations (e.g., \citealt{PineaudesForets+etal+1992,LeBourlot+etal+1993}), and we plan to clarify this phenomenon in a future work due to its complexity.

\begin{figure} 
  \centering
  \includegraphics[width=14.0cm, angle=0]{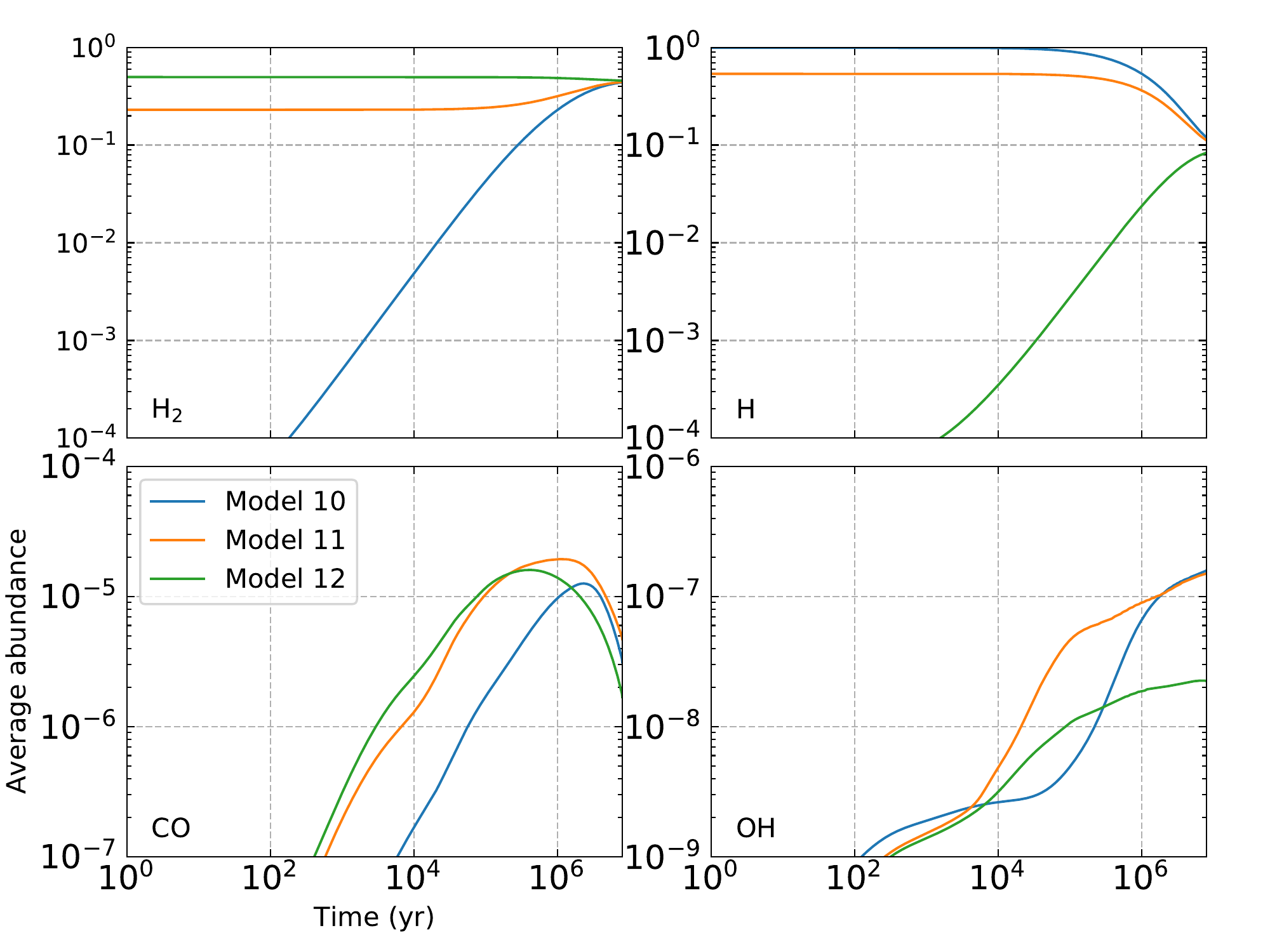}
  \caption{The average abundances for H, H$_2$, CO, and OH as a function of time using models 10-12.
     } 
  \label{FigAverageAbundanceForStatic}
\end{figure}
Fig~\ref{FigAverageAbundanceForStatic} shows the average abundances for above four species varying with time in models 10-12.
For models 10-11, the CO and OH average abundances at $t\sim10^6$ yr reach a high level, when the HI column density of $\sim8{\times}10^{21}$ cm$^{-2}$ cannot match what has been observed (e.g., \citealt{Wannier+etal+1983,Bigiel+etal+2008,Lee+etal+2012}).
The CO abundance for late time reaches a low level of $4{\times}10^{-6}$ due to CO freeze out, when the OH abundance of $\sim2{\times}10^{-7}$ is growing, so the CO-poor cloud may be traced by OH.
It reveals that a molecular cloud with typical physical parameters cannot stay static for a long time and still be observable in CO.
Accordingly, these static clouds in which hydrogen are mainly atomic initially cannot match observational characteristics, unless we adopt the argument from \cite{Seifried+etal+2022} that the underestimation of HI column density is due to both the large HI temperature variations and the effect of noise in regions of high optical depth, and real HI column densities of $\gtrsim10^{22}$ cm$^{-2}$ are frequently reached in molecular clouds. 
In Model 12, the HI average abundance for $t>10^6$ yr can match their observed values in molecular clouds, while the OH abundance for all time is lower than its observed value of $5{\times}10^{-8}$ (\citealt{Crutcher+1979}).
Hence the static cloud in which hydrogen is molecular initially also cannot match OH observed abundance in typical molecular clouds.
In short, the static clouds cannot match observational characteristics in typical molecular clouds, which is one main reason why we consider collapsing models.

\section{The influence of some physical parameters on chemical evolution}
\label{Sec:influence of physical parameters}
In previous sections we have assumed constant value for some physical parameters ($T_{\rm gas}$, $T_{\rm dust}$, $G_0$, and $\zeta$).  
Here we study whether adopting other values for them might affect chemical evolution, using the static models as an example.
To obtain appropriate values of physical parameters, we analyze the influences of above four physical parameters on H$_2$, being the most important molecule in clouds.

\subsection{Insensitivity to gas temperature}
\label{subsect:Tgas}
The relationship between the H adsorption rate coefficient $k_{\rm H}$ and gas temperature is given by $k_{\rm H} \varpropto T_{\rm gas}^{1/2}$ (\citealt{Du+2021}), so H$_2$ formation is accelerated with increasing gas temperature.
We select a high gas temperature of 100 K to study the changes in H, H$_2$, CO, and OH average abundances.
We find the average abundances of H, CO, and OH do not change appreciably, relative to the case with $T_{\rm gas}=30$ K.
Namely, chemical evolution is insensitive to gas temperature when it is lower than 100 K.

\subsection{Dependence on dust temperature}
\label{subsect:Tdust}
\begin{figure} 
  \centering
  \includegraphics[width=14.0cm, angle=0]{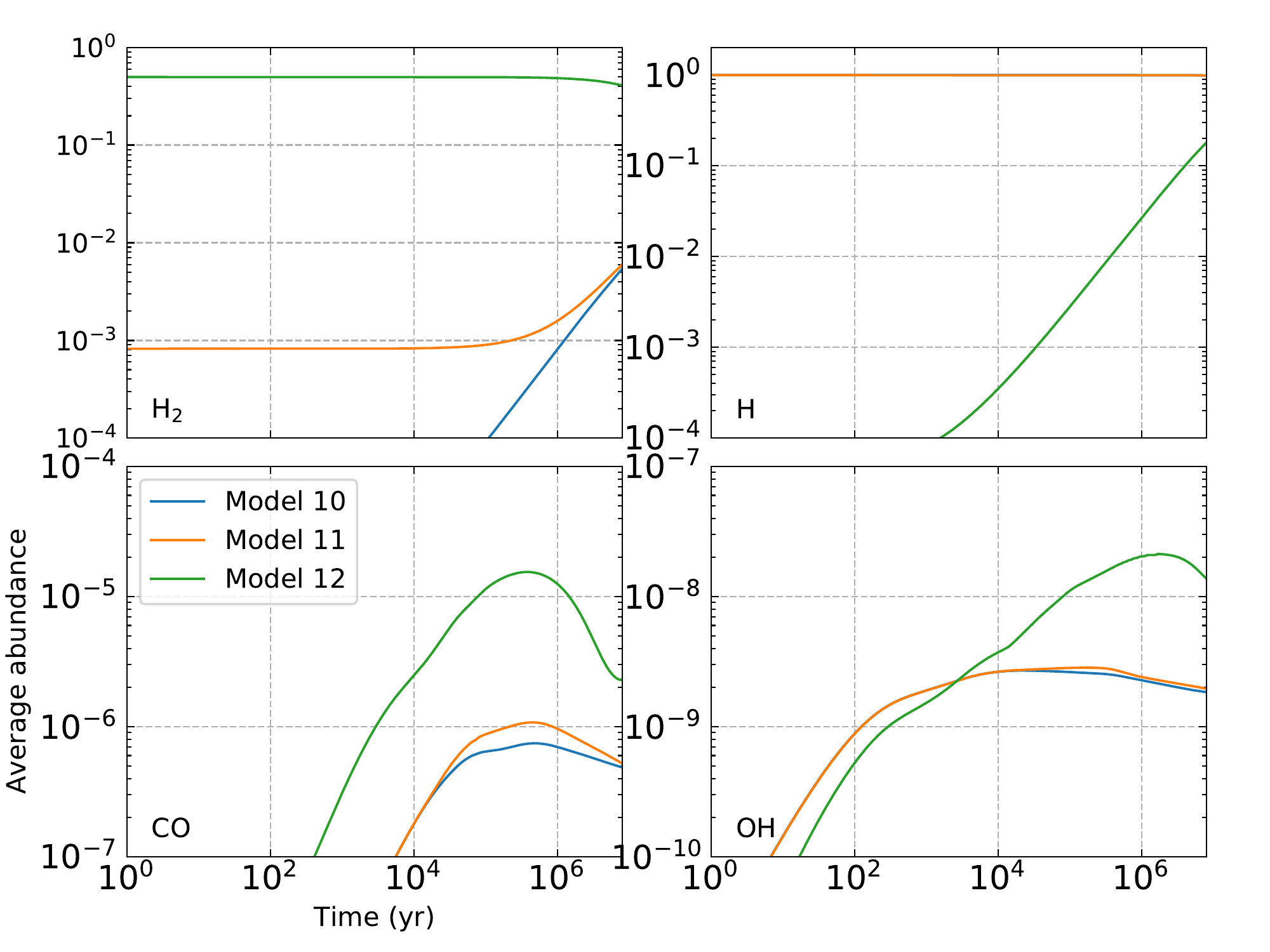}
  \caption{The average abundances for H, H$_2$, CO, and OH as a function of time using models 10-12 with $T_{\rm dust}=20$ K.
     } 
  \label{FigAverageAbundanceAtHighTdust}
\end{figure}
The competition between desorption and diffusion rate of H atoms on dust grain surfaces leads to great differences in H$_2$ formation efficiency at different dust temperature.
The H thermal desorption rate dominating its evaporation is given by 
\begin{equation}
  k_{\rm des} = 2.74{\times}10^{12}~{\rm s}^{-1} {\rm exp}(-450 {\rm K}/T_{\rm dust}),
\end{equation} 
and the diffusion rate for H atoms $k_{\rm diff}$ is 
\begin{equation}
  k_{\rm diff} = 2.1{\times}10^6~{\rm s}^{-1} {\rm exp}(-225 {\rm K}/T_{\rm dust}),
\end{equation}
obtained using the H desorption energy of 450 K, a rectangular barrier with a height of 225 K, and the number of sites on a grain surface of $1.3{\times}10^6$ (\citealt{Du+2021}).
When the H desorption rate equals its diffusion rate, the dust temperature is close to $\sim16$ K.
The H atoms can scan all of the adsorption sites before desorption at dust temperature below $\sim16$ K, so the H$_2$ formation rate is determined by the H adsorption rate (independent of dust temperature).
When the dust temperature is higher than $\sim16$ K, considerable H atoms go back to the gas phase before meeting their reaction partners, resulting the H$_2$ formation efficiency is significantly lower than unity.
Hence we select a high dust temperature of 20 K to study how it affects chemical evolution in Fig~\ref{FigAverageAbundanceAtHighTdust}.

At $T_{\rm dust}=20$ K, the static clouds at late time using models 10-12 is far from reaching chemical equilibrium. 
This is mainly because the desorption rates for some species on dust grain surfaces are boosted with increasing dust temperature, resulting in some surface reactions become inefficient.
In Model 10-11, the abundances of H$_2$, CO and OH are different from in Fig~\ref{FigAverageAbundanceForStatic}, which suggests some important molecules formed on dust grain surfaces heavily depend on the dust temperature.
In Model 12, the average abundances for these four species at $T_{\rm dust}=20$ K seem to not change appreciably relative to Fig~\ref{FigAverageAbundanceForStatic} with $T_{\rm dust}=15$ K.
Therefore, CO clouds are difficult to form at high dust temperatures, which is one reason why most observed dust temperatures in molecular clouds are lower than 20 K (e.g., \citealt{Bianchi+etal+2003,Kramer+etal+2003,Schnee+Goodman+2005}). 
In general, the static clouds with a high dust temperature cannot match observational characteristics.

\subsection{Variation with the intensity of interstellar radiation filed}
\label{subsect:G0}
\begin{figure} 
  \centering
  \includegraphics[width=14.0cm, angle=0]{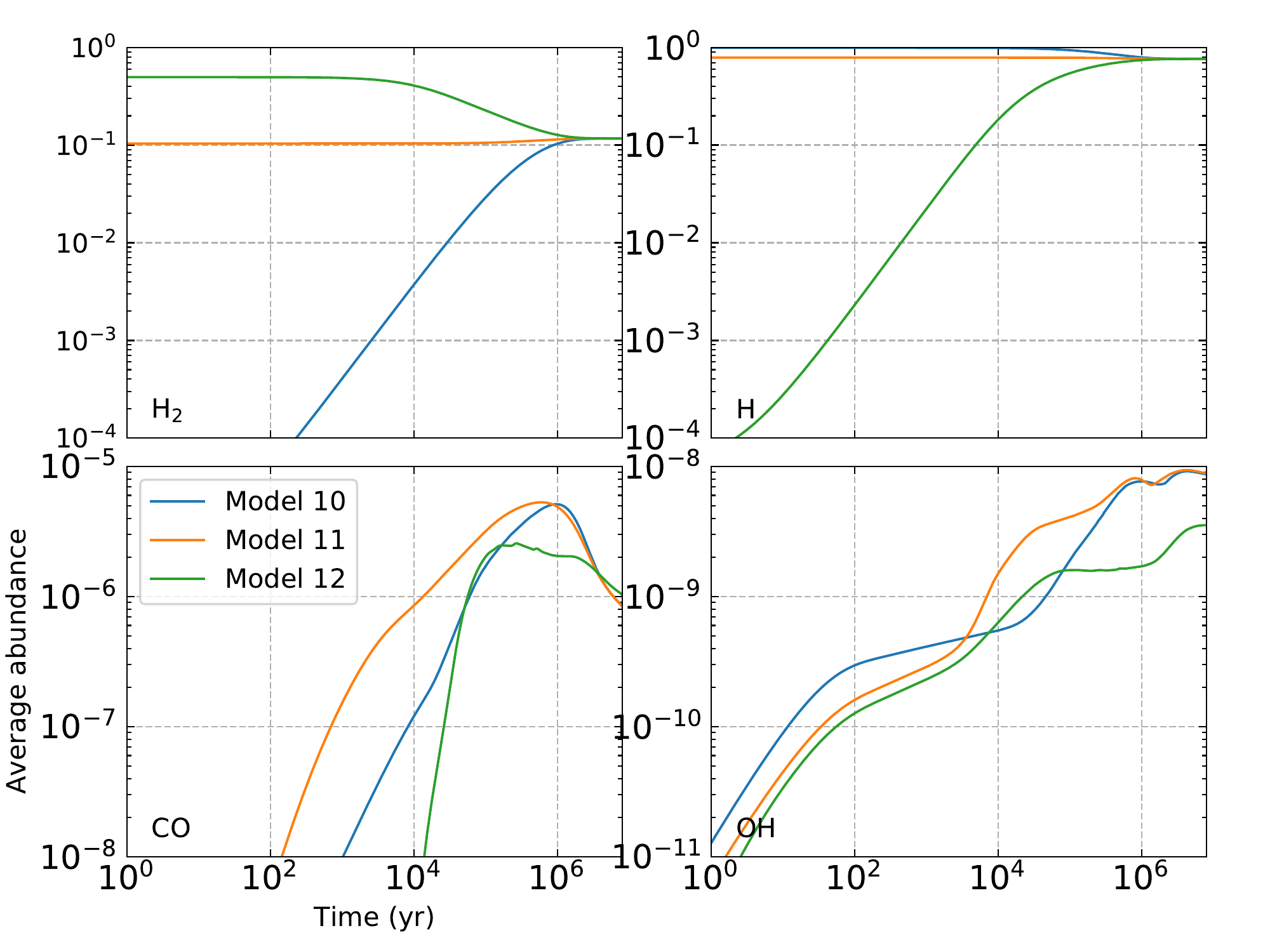}
  \caption{The average abundances for H, H$_2$, CO, and OH as a function of time using models 10-12 with $G_0=1000$. The intensity scaling factor $G_0$ is defined by the actual radiation field relative to \cite{Draine+1978} radiation field.
     } 
  \label{FigAverageAbundanceWithHighG0}
\end{figure}

The H$_2$ photodissociation rate is closely correlated with the H$_2$ self-shielding and shielding by dust (\citealt{Draine+Bertoldi+1996}), so the effect of interstellar radiation filed (ISRF) on H$_2$ is mainly limited to the outer layers of clouds.
The intensity of radiation field ($G_0$) near a star can easily exceed 1000.
Therefore, we adapt models 10-12 with $G_0=1000$ in Fig~\ref{FigAverageAbundanceWithHighG0} to study chemical evolution.
 
In models 10-12 with a strong ISRF, the abundances of H$_2$, CO and OH for all time are much lower relative to Fig~\ref{FigAverageAbundanceForStatic} with a weak ISRF.
At $t\sim10^6$ yr, the CO abundances in these models are close to $4{\times}10^{-6}$, so these static clouds are CO-poor.
The CO molecules in molecular clouds exposed to strong ISRF can only exist in large quantities for a short time, which is one reason why molecular clouds near early-type stars become quickly undetectable in CO (e.g., \citealt{Hartmann+etal+2001}).
At this time, these static clouds are mainly atomic, and their HI column density of $\sim10^{22}$ cm$^{-2}$ exceed observed upper threshold of a few $10^{21}$ cm$^{-2}$.
In general, the static clouds with a strong ISRF cannot match observational characteristics in typical molecular clouds. 

\subsection{Variation with cosmic-ray ionization rate}
\label{subsect:CRIR}
\begin{figure} 
  \centering
  \includegraphics[width=14.0cm, angle=0]{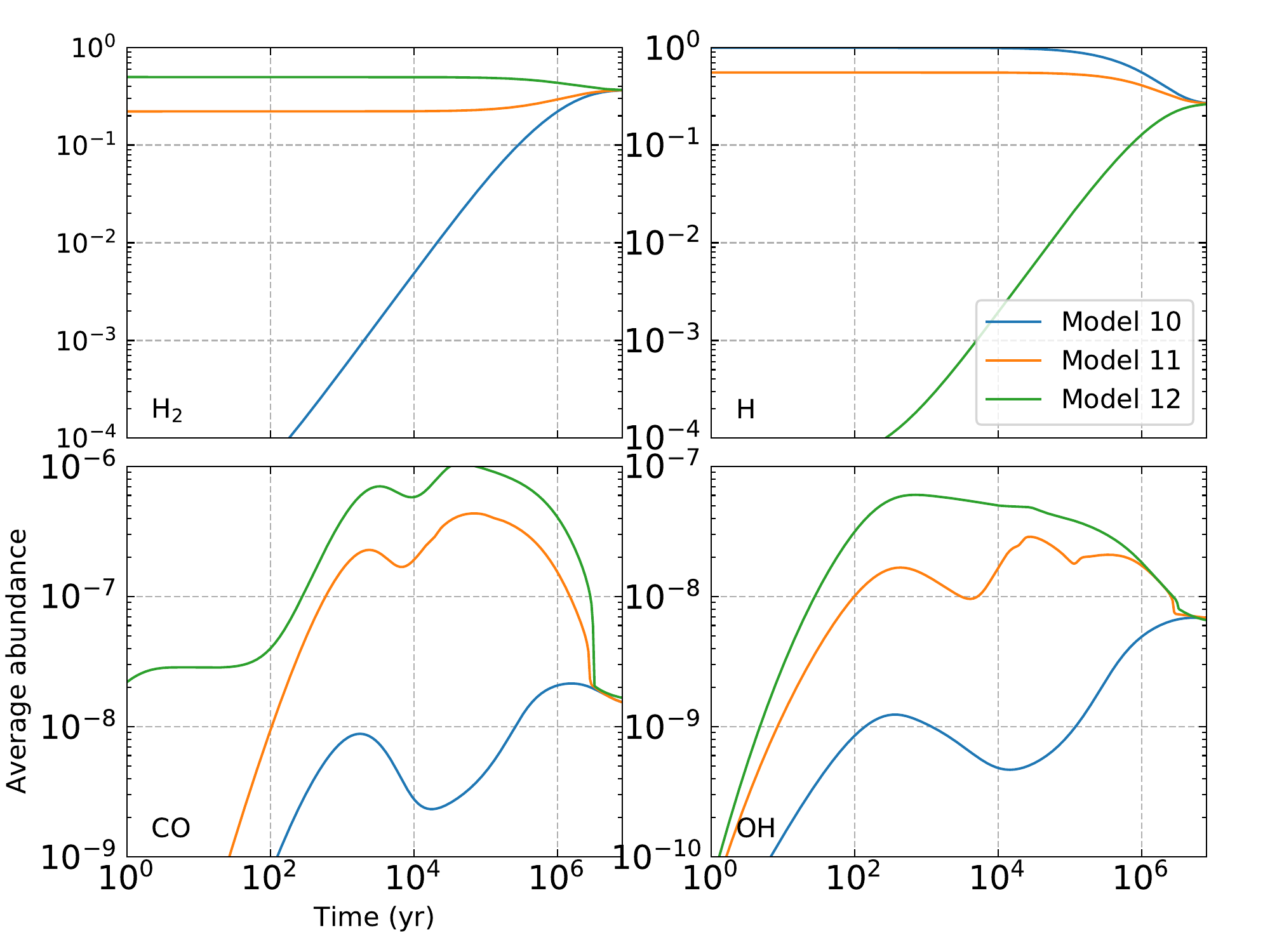}
  \caption{The average abundances for H, H$_2$, CO, and OH as a function of time using models 10-12 with a cosmic-ray ionization rate of $3{\times}10^{-15}$ s$^{-1}$.
     } 
  \label{FigAverageAbundanceWithHighChi}
\end{figure}

The observed cosmic-ray ionization rates as traced by H$_3^+$ are mostly in a narrow range in the Galactic diffuse interstellar medium and molecular clouds, from a few times $10^{-17}$ s$^{-1}$ to a few times $10^{-16}$ s$^{-1}$ (e.g., \citealt{McCall+etal+1999, Indriolo+McCall+2012}).
The cosmic-ray ionization rates near Supernova Remnant IC 443 and toward the Galactic center are in a range, from a few times $10^{-16}$ s$^{-1}$ to a few times $10^{-14}$ s$^{-1}$ (e.g., \citealt{Hartquist+etal+1978,Indriolo+etal+2010,Geballe+Oka+2010}).
Accordingly, the cosmic-ray ionization rates in different interstellar medium have a great difference.
The attenuation of cosmic rays is negligible (\citealt{Umebayashi+Nakano+1981}) at column density below $4{\times}10^{25}$ cm$^{-2}$, so intensive cosmic rays greatly influence the molecules formation for all positions.
Thus we adopt a high cosmic-ray ionization rate ($\zeta$) of $3{\times}10^{-15}$ s$^{-1}$ to study chemical evolution, and the results are shown in Fig~\ref{FigAverageAbundanceWithHighChi}.

When the cosmic-ray ionization rate is adopted as $3{\times}10^{-15}$ s$^{-1}$, these static clouds at late time are mainly composed of molecular hydrogen, but the CO and OH abundances are extremely low.
The observation for CO-dark molecular gas indicates that OH column density ranges from $10^{12}$ to $10^{14}$ cm$^{-2}$, that the upper limit of CO column density is $5{\times}10^{14}$ cm$^{-2}$ at a CO sensitivity of 0.25 K, and that HI column density is from $10^{20}$ to $10^{21}$ cm$^{-2}$ (\citealt{Li+etal+2018}).
Hence these static clouds for late time where CO column density is close to $\sim4{\times}10^{14}$ cm$^{-2}$ may be CO-dark, whereas they have a high OH column density of $\sim10^{14}$ cm$^{-2}$ and may be detectable with OH absorption. 
In Model 12, the abundances of CO and OH basically reach their respective maximum of $10^{-6}$ and $6{\times}10^{-8}$ at $t\sim10^4$ yr, then the static cloud mainly composed of molecular hydrogen quickly becomes transparent in CO and OH.
Accordingly, CO and OH molecules in clouds with intensive cosmic rays cannot exist in large quantities for a long time, unless in some high density regions.
In general, the static clouds with intensive cosmic rays cannot match observational characteristics in typical molecular clouds.
\section{Discussions}
\label{sect:discussions} 
The treatments of initial abundances are simple in many chemical simulations (e.g., \citealt{Hasegawa+etal+1992,Garrod+Pauly+2011}), but initial chemical compositions can greatly affect the chemical evolution.
Therefore, we study three types of initial abundances for each types of clouds (collapsing clouds, and static clouds) in Sect~\ref{sect:results}. 
The chemical evolution in these clouds show that the models in which hydrogen is mainly atomic initially (e.g., Model 1) seem excel the models in which hydrogen is mainly molecular initially (e.g., Model 3) in matching the observational characteristics in molecular clouds.
Accordingly, we suggest that initial chemical compositions of molecular clouds is mainly atomic hydrogen.
This means CO molecules are able to form even without the pre-existence of H$_2$.

The dynamics of clouds is an important factor affecting the chemical evolution of clouds.
Thus we show the chemical evolution (e.g., H, H$_2$, CO and OH) in three types of clouds with different dynamics in Sect~\ref{sect:results}.
Chemical evolution in these clouds with identical initial chemical compositions (e.g., models 1 and 10) show that the collapsing models are better than the static models in matching the observational characteristics in molecular clouds.
This is mainly because gravity can enhance the reaction rates by increasing the number density during collapse, then speeds up the formation of some important molecules (e.g., H$_2$, CO, OH).
The differences in morphology between prolate, oblate, and spherical clouds lead to large differences in chemical evolution, via some differences in the evolution of number density (see Sect~\ref{sect:EllipsoidalCloudModel}).

In Sect~\ref{Sec:influence of physical parameters}, we explore whether adopting other values for some important physical parameters might affect the conclusions in static models.
Some important molecules heavily depend on the dust temperature, so the heating and cooling mechanisms should be taken into consideration, to automatically solve the temperature in tandem with chemical abundances (e.g., as done in \cite{Du+Bergin+2014}).
Under the circumstances of high dust temperature, strong radiation field, and intensive cosmic rays; the static models seem unable to match the observational characteristics in molecular clouds.
Namely, some important molecules (e.g., H$_2$, CO) in clouds are easier to form under typical molecular clouds environment.

The uncertainties in the magnetic fields strength and the turbulence, together with a possible revision of the lifetimes of the molecular gas (e.g., \citealt{Hartmann+etal+2001}), have resulted in two opposed views of clouds lifetime, including a long cloud lifetime of at least 10 Myr (e.g., \citealt{Mouschovias+etal+2006,Schinnerer+etal+2013}) and a short cloud lifetime of 3-5 Myr (e.g., \citealt{Vazquez-Semadeni+etal+2003,Hartmann+2003}).
From a chemical modeling perspective, the CO cloud lifetime estimates do not include any earlier phase when the gas is dominantly molecular but CO has not formed in sufficient quantity for detectable emission (\citealt{Bergin+Tafalla+2007}).
In our models in which hydrogen is mainly atomic initially, above earlier phases do not seem to exist, so CO cloud lifetime estimates are not affected.
At high gas temperature, strong radiation field, and intensive cosmic rays regions as in Sect~\ref{Sec:influence of physical parameters}; CO clouds in such environments are transient due to rapid CO destruction (related to CO adsorption, photodissociation, and cosmic-ray-induced photodissociation reactions).

\section{Conclusions}
\label{sect:conclusions}
In this paper, we model chemical evolution for collapsing spherical, collapsing ellipsoidal clouds, and static clouds using three types of initial abundances.
We also study the influence of gas temperature, dust temperature, intensity of interstellar radiation field, and cosmic-ray ionization rate on chemical evolution in static clouds.
Our conclusions are as follows:\\
1) The collapsing models are better than the static models in matching the observational characteristics in typical molecular clouds.
This is mainly because gravity can speed up the formation of some important molecules (e.g., H$_2$, CO, OH) by increasing the number density during clouds collapse.\\
2) The differences in morphology between prolate, oblate, and spherical clouds can lead to differences in chemical evolution, which are mainly due to their different evolution of number density.\\
3) In the framework of our modeling, H atoms can accelerate OH formation by two major reactions: O + H $\rightarrow$ OH in gas phase and on dust grain surfaces, as a consequence the models in which hydrogen is mainly atomic initially perform better than the models in which hydrogen is mainly molecular initially in matching observations. 
Therefore, we suggest the initial chemical compositions of molecular clouds to be mainly atomic hydrogen.\\
4) Under the circumstances of high gas temperature, strong radiation field, and intensive cosmic rays, the static CO clouds are transient due to rapid CO destruction.

\section*{Acknowledgements}
This work is financially supported by the National Science Foundation of China through grants 12041305 and 11873094.

\bibliographystyle{raa}
\bibliography{bibtex}

\appendix
\section{Figures in collapsing ellipsoidal clouds}
\label{Figures}
\begin{figure} 
  \centering
  \includegraphics[width=14.0cm, angle=0]{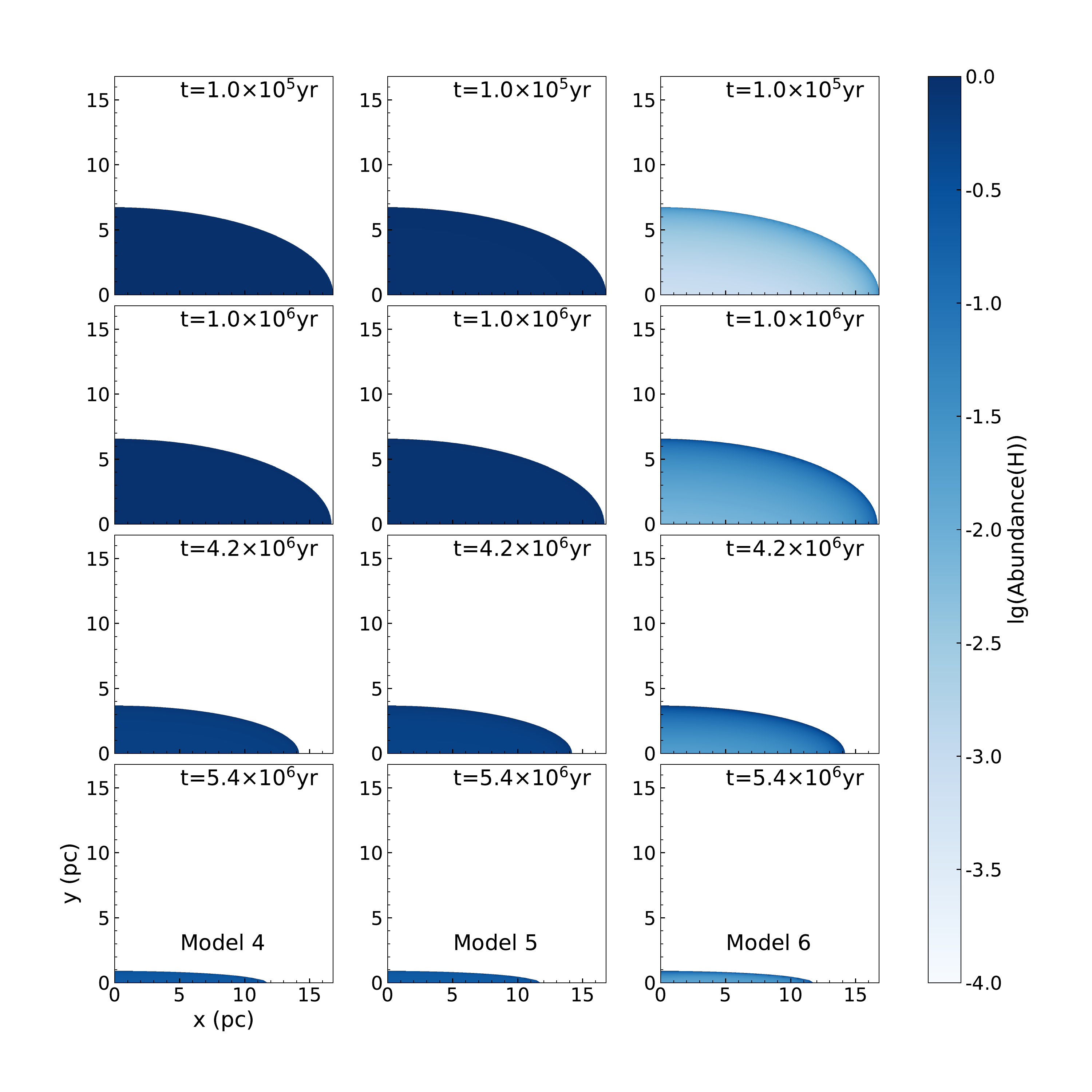}
  \caption{The H abundance distribution at different time in a gravitationally collapsing uniform oblate ellipsoidal cloud using models 4-6.
  Note that as time goes on, the cloud shrinks.} 
  \label{FigAtomicHydrogenAbundanceForOblateCase}
\end{figure}

\begin{figure} 
  \centering
  \includegraphics[width=14.0cm, angle=0]{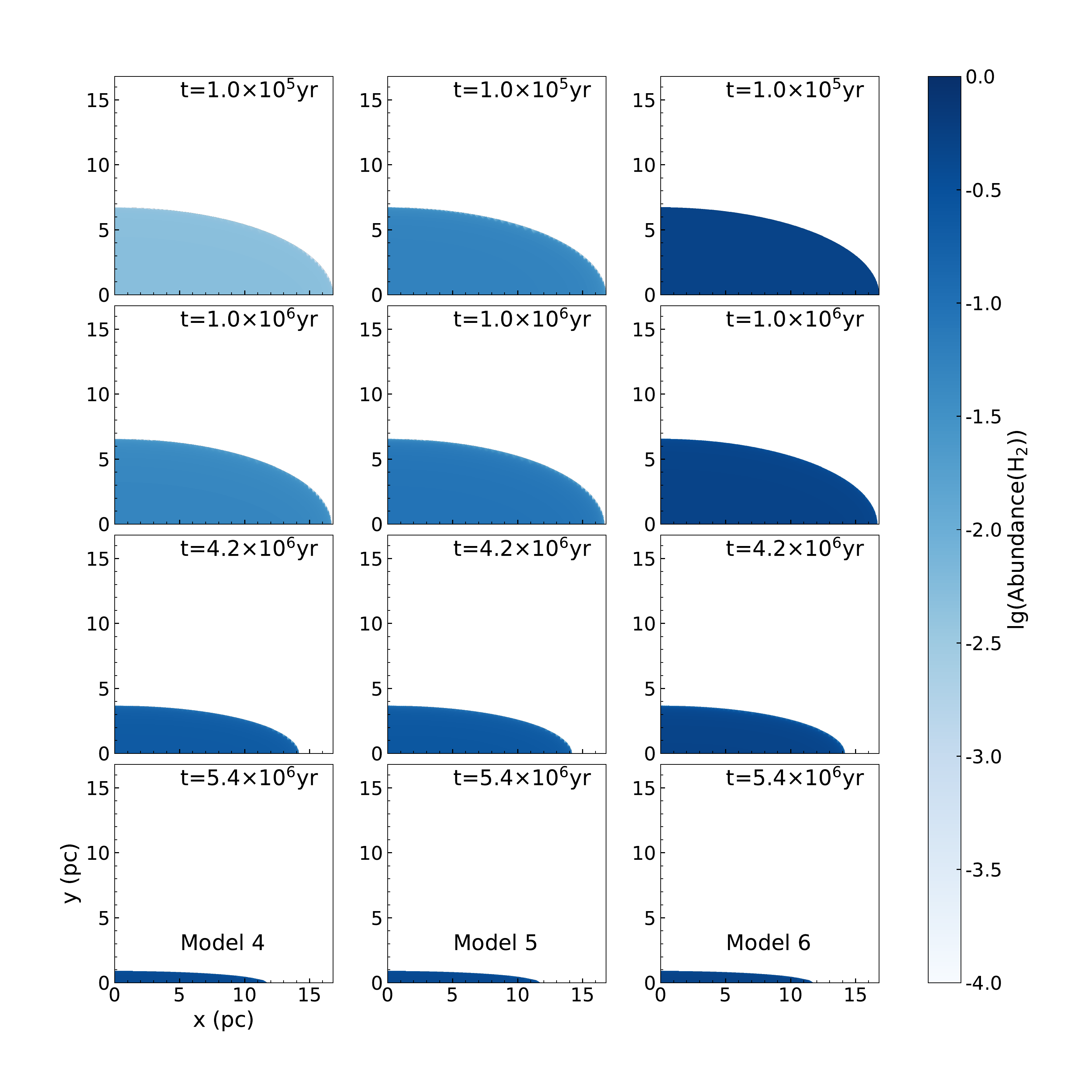}
  \caption{The H$_2$ abundance distribution at different time in models 4-6.}
  \label{FigMolecularHydrogenAbundanceForOblateCase}
\end{figure}

\begin{figure} 
  \centering
  \includegraphics[width=14.0cm, angle=0]{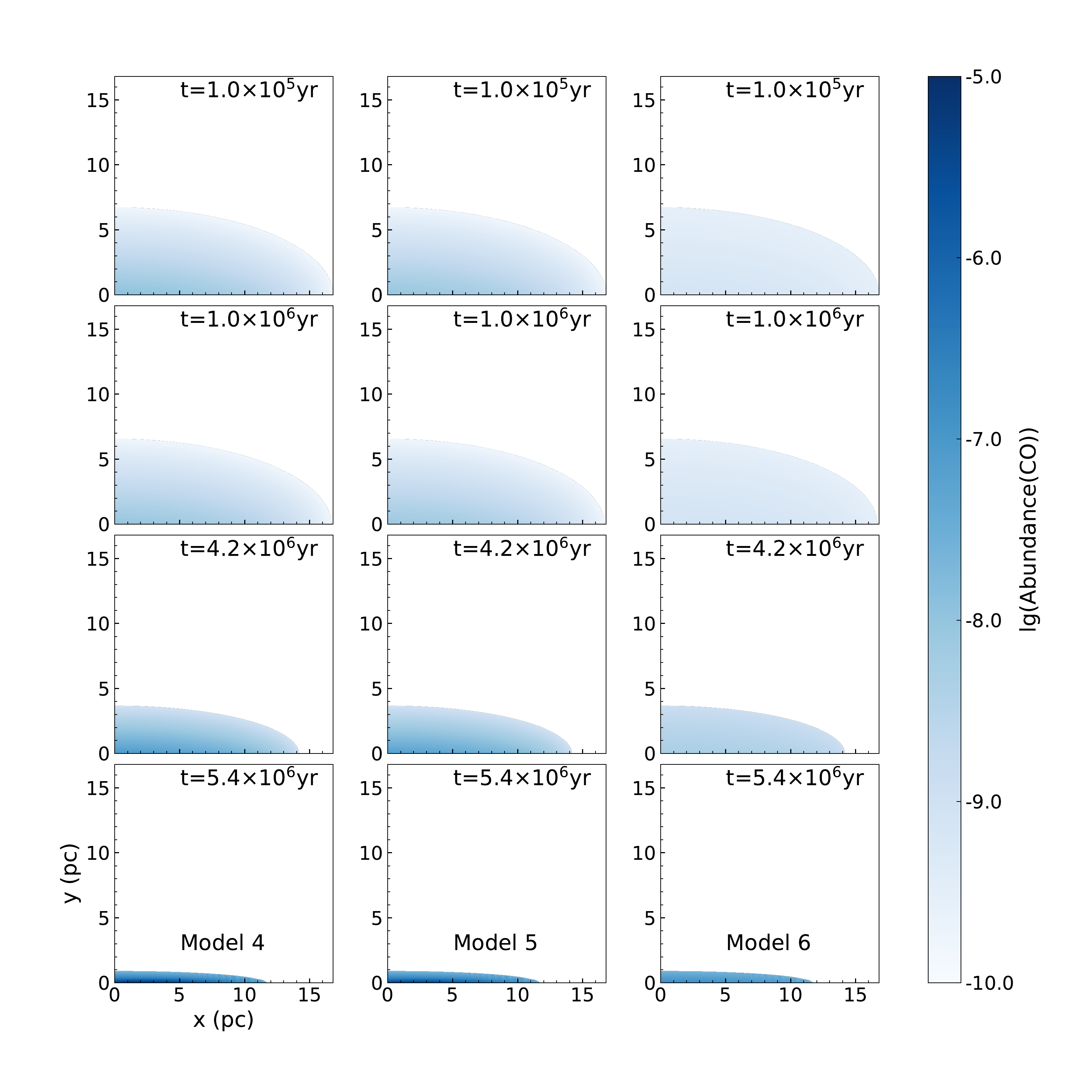}
  \caption{The CO abundance distribution at different time in models 4-6.}
  \label{FigCOAbundanceForOblateCase}
\end{figure}

\begin{figure} 
  \centering
  \includegraphics[width=14.0cm, angle=0]{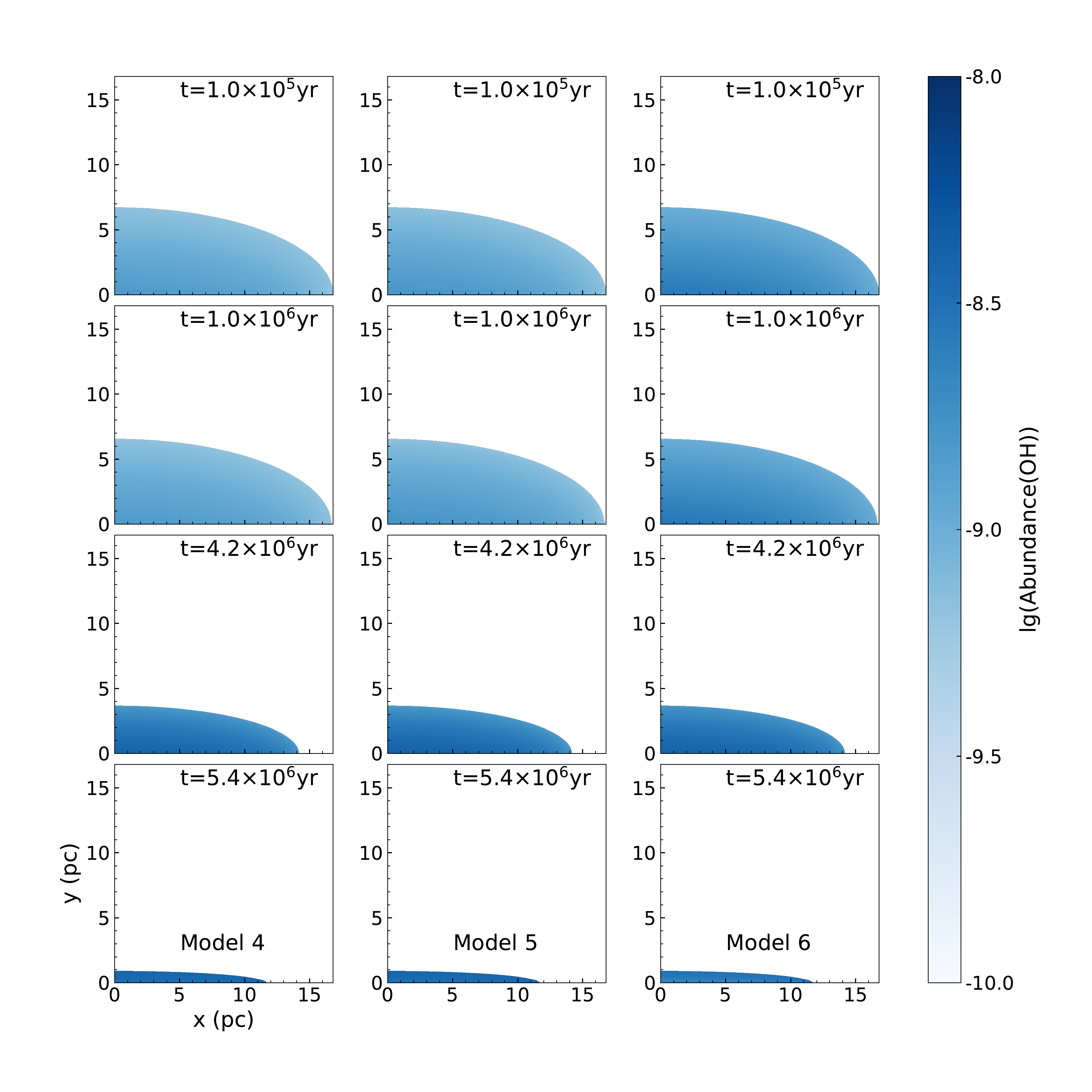}
  \caption{The OH abundance distribution at different time in models 4-6.}
  \label{FigOHAbundanceForOblateCase}
\end{figure}

\begin{figure} 
  \centering
  \includegraphics[width=14.0cm, angle=0]{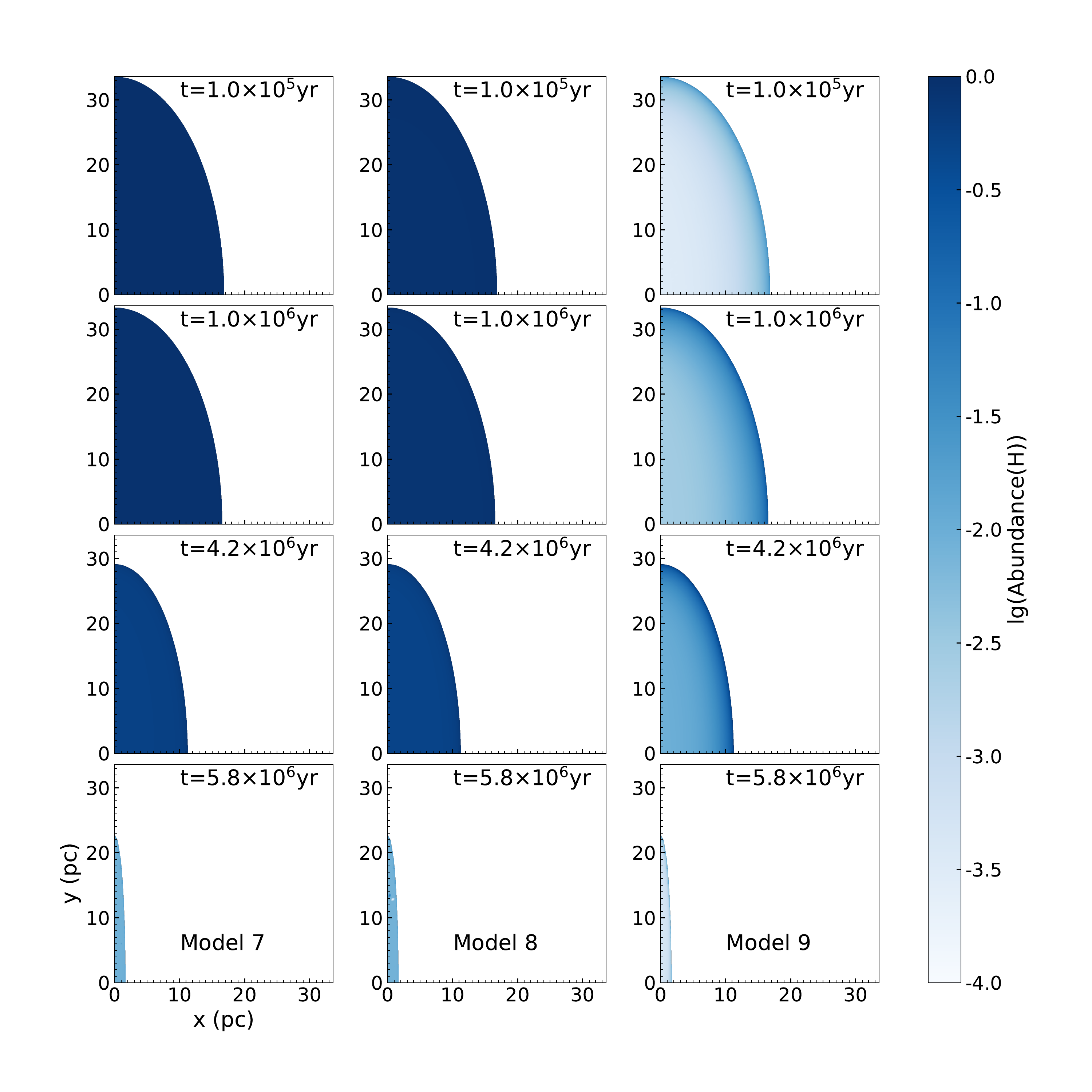}
  \caption{The H abundance distribution at different time in a gravitationally collapsing uniform prolate ellipsoidal cloud using models 7-9.} 
  \label{FigAtomicHydrogenAbundanceForProlateCase}
\end{figure}

\begin{figure} 
  \centering
  \includegraphics[width=14.0cm, angle=0]{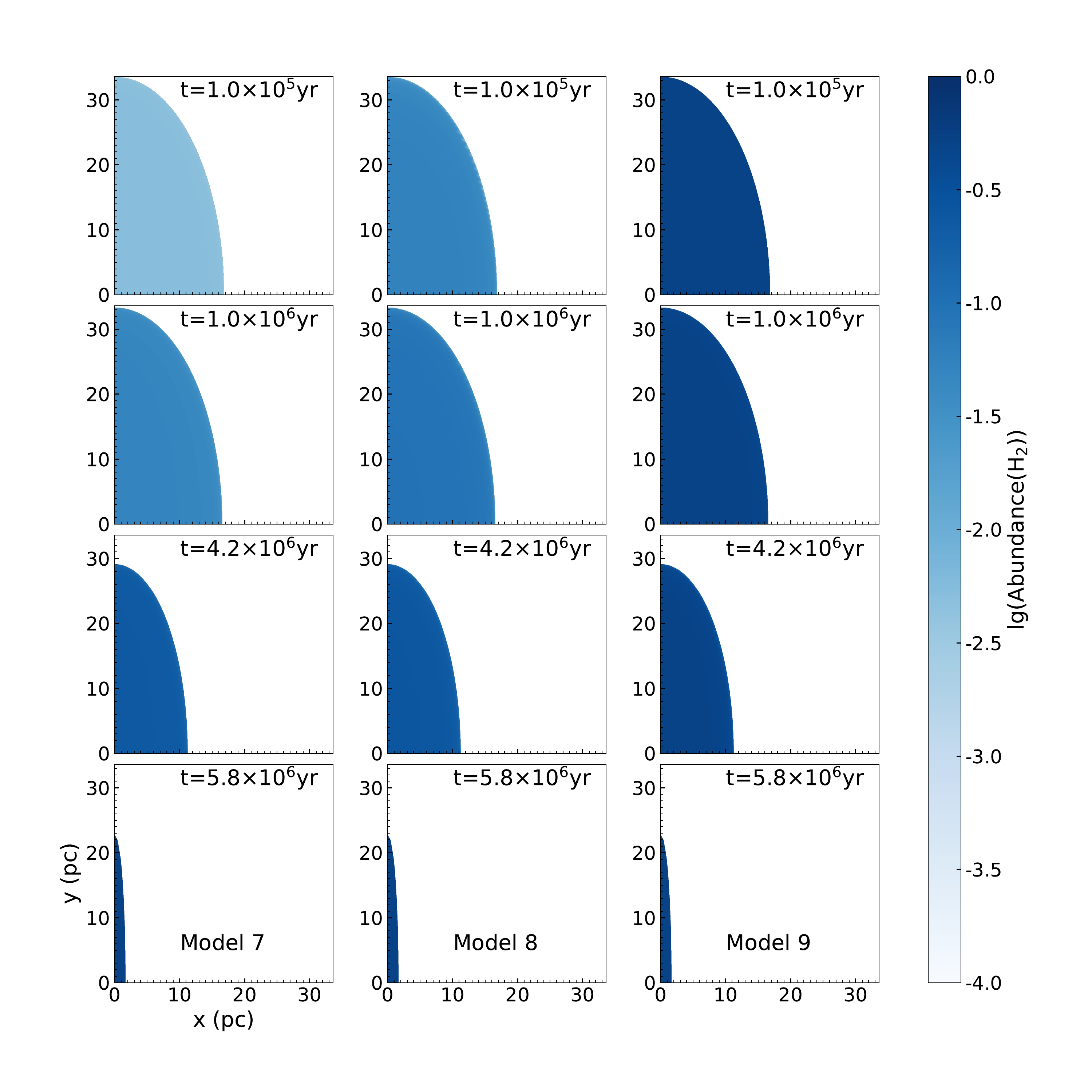}
  \caption{The H$_2$ abundance distribution at different time in models 7-9.}
  \label{FigMolecularHydrogenAbundanceForProlateCase}
\end{figure}

\begin{figure} 
  \centering
  \includegraphics[width=14.0cm, angle=0]{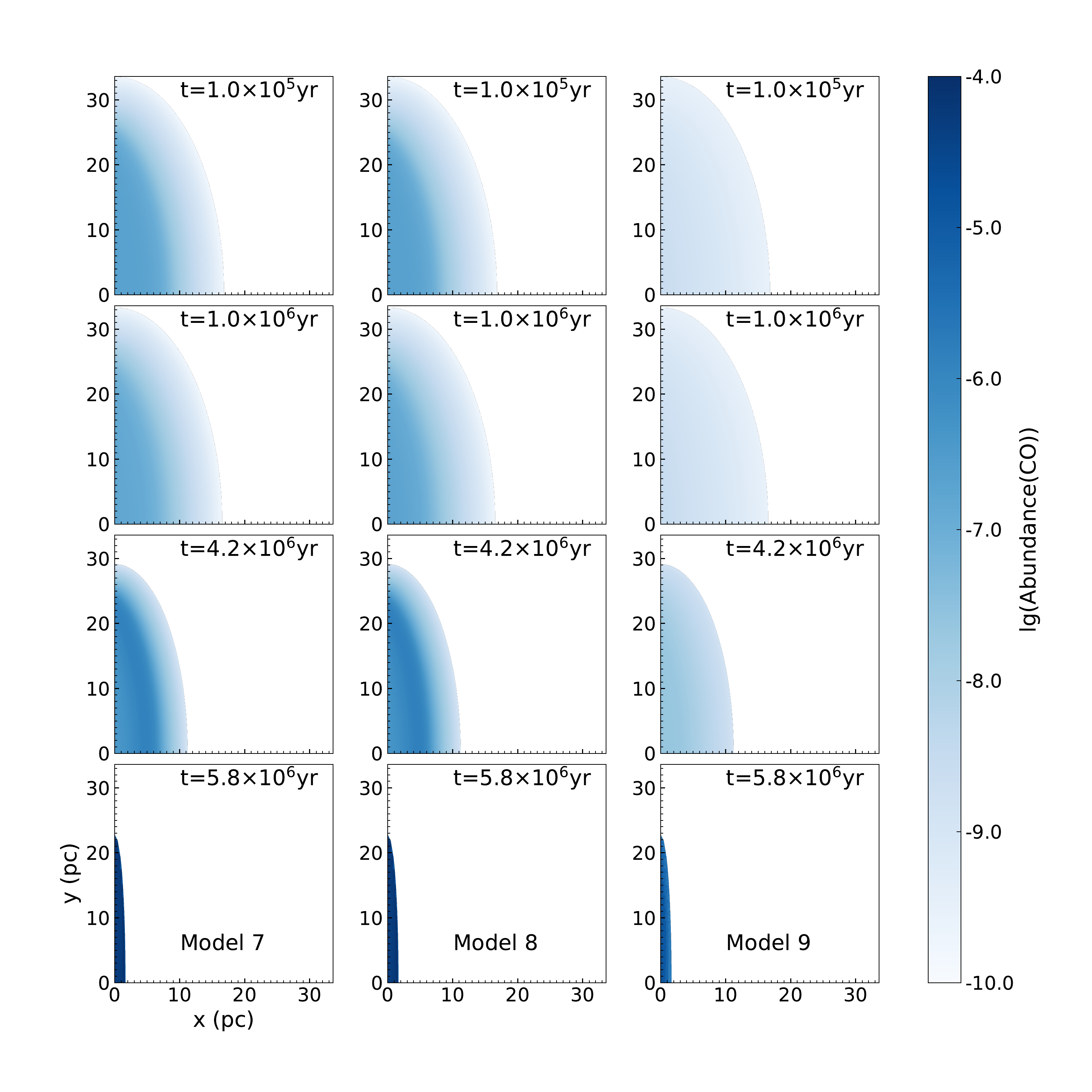}
  \caption{The CO abundance distribution at different time in models 7-9.}
  \label{FigCOAbundanceForProlateCase}
\end{figure}

\begin{figure} 
  \centering
  \includegraphics[width=14.0cm, angle=0]{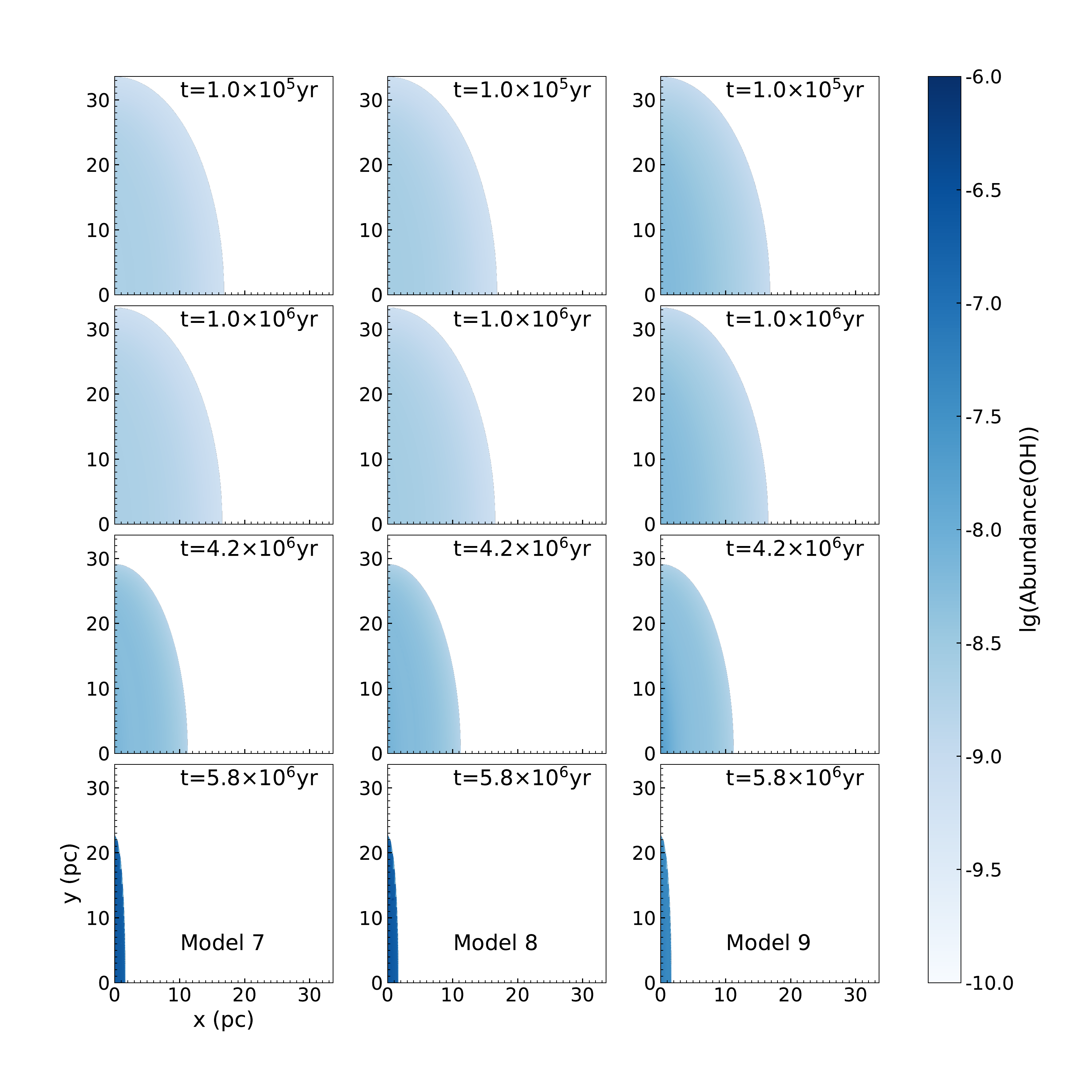}
  \caption{The OH abundance distribution at different time in models 7-9.}
  \label{FigOHAbundanceForProlateCase}
\end{figure}

\end{document}